\UseRawInputEncoding
\documentclass[twocolumn,apl,epsf,superscriptaddress]{revtex4-2}

\usepackage{amsmath}
\usepackage{amsfonts}
\usepackage{amssymb}
\usepackage{epsfig}
\usepackage{epsf}
\usepackage{array}
\usepackage{color}
\usepackage{ulem}
\usepackage{units}

\newcommand{\vect}[1]{\mathbf #1}

\usepackage{setspace}
\begin{document}

%\title{Study of Oxide Interfaces Using Model Hamiltonians: Past, Present, and Future} 
\title{Novel phenomena in transition-metal oxide thin films and heterostructures with strong correlations and spin-orbit coupling}
\altaffiliation{Copyright  notice: This  manuscript  has  been  authored  by  UT-Battelle, LLC under Contract No. DE-AC05-00OR22725 with the U.S.  Department  of  Energy.   
The  United  States  Government  retains  and  the  publisher,  by  accepting  the  article  for  publication, 
acknowledges  that  the  United  States  Government  retains  a  non-exclusive, paid-up, irrevocable, world-wide license to publish 
or reproduce the published form of this manuscript, or allow others to do so, for United States Government purposes.  
The Department of Energy will provide public access to these results of federally sponsored  research  in  accordance  with  the  DOE  Public  Access  Plan 
(http://energy.gov/downloads/doe-public-access-plan)}. 
%
%%%
\author{Satoshi Okamoto}
\altaffiliation{okapon@ornl.gov}
\affiliation{Materials Science and Technology Division, Oak Ridge National Laboratory, Oak Ridge, Tennessee 37831, USA}
\author{Narayan Mohanta}
\affiliation{Materials Science and Technology Division, Oak Ridge National Laboratory, Oak Ridge, Tennessee 37831, USA}
\affiliation{Department of Physics, Indian Institute of Technology Roorkee, Roorkee 247667, India}
\author{Ho Nyung Lee}
\affiliation{Materials Science and Technology Division, Oak Ridge National Laboratory, Oak Ridge, Tennessee 37831, USA}
\author{Adriana Moreo}
\author{Elbio Dagotto}
\affiliation{Department of Physics and Astronomy, The University of Tennessee, Knoxville, Tennessee 37996, USA}
\affiliation{Materials Science and Technology Division, Oak Ridge National Laboratory, Oak Ridge, Tennessee 37831, USA}

%%%%%%%%%%%%%%%%%%%%%%%%%%%%%%%%%%%%%%%%%%%%%%%%%%%%%%%%%%%%%%%%%%
\begin{abstract}
Transition-metal oxides have been a central subject of condensed matter physics for decades. 
In addition to novel electronic states driven by the influence of strong correlation, 
relativistic spin-orbit coupling effects have recently attracted much 
attention for their potential to explore topological phenomena. 
In this article, we review various experimental and theoretical studies on transition-metal oxides with
focus on thin films and heterostructures where their physics is much influenced by 
correlation effects and spin-orbit coupling. The combination of the heterostructure geometry together
with correlation and topology leads to a variety of novel states here reviewed. 
We also discuss perspectives for future research in this broad promising area.
\end{abstract}
\maketitle
%%%%%%%%%%%%%%%%%%%%%%%%%%%%%%%%%%%%%%%%%%%%%%%%%%%%%%%%%%%%%%%%%%%%%

\section{Introduction}

Bulk transition-metal oxides (TMOs) have played a major role in exploring novel electronic behaviors arising from strong correlations, 
such as high-$T_c$​ superconductivity in cuprates and the colossal magnetoresistance (CMR) effect in manganites \cite{Imada1998,Tokura2000}. 
Strongly correlated systems are those where the electronic repulsion between electrons is larger, or
at least comparable, to the strength of the kinetic energy portion of the Hamiltonian. 
In the language of the famous Hubbard model 
that means $U$ larger than $W$, where $U$ is the on-site repulsion and $W$ the bandwidth of the bands induced by the
hopping portion of the Hamiltonian. In strongly correlated systems, we cannot start a theoretical analysis assuming a dominant hopping portion, because we are in the opposite limit $U$ larger than $W$. Thus, in this regime we have
to resort to techniques other than perturbation theory in $U$, 
such as expansions in $W/U$, variational and mean-field methodologies,
or a wide array of computational techniques in finite clusters.

With advancements in thin-film growth techniques, the study of correlated TMOs has entered a new stage. 
By tuning the film thickness, combining different materials, and introducing strain fields from substrates, 
researchers can modify the many-body electronic states of TMOs \cite{Ohtomo2002,Okamoto2004}. 
This approach has been widely used to control the ferroelectric behavior of titanate heterostructures \cite{Lee2005}, 
as well as the superconducting and magnetic behaviors of manganite/cuprate heterostructures \cite{Sefrioui2003,Chakhalian2006,Nemes2008,Salafranca2010,Okamoto2010}. 
Additionally, it has enabled the study of interfacial magnetism and orbital 
reconstruction in BiFeO$_3$/(La,Sr)MnO$_3$  \cite{Yu2010a}, 
and charge transfer and magnetism in iridate/manganite  \cite{Nichols2016,Okamoto2017} heterostructures.
Clearly, at the interface of two correlated systems, one system may influence on the other
and {\it vice versa}. Thus, novel states can be found.

Magnetism influenced by spin-orbit coupling (SOC) appears in many contexts. 
In systems with broken inversion ($\cal P$) symmetry, which is relevant to TMO heterostructures, 
an antisymmetric interaction known as the Dzyaloshinskii-Moriya (DM) interaction (DMI) \cite{Dzyaloshinskii1957,Moriya1960} 
emerges as an interplay between correlation effects and SOC. 
This interaction is expressed as
\begin{equation}
{\cal H}_{\rm DMI} = -\sum_{\langle ij \rangle}\vect D_{ij} \cdot (\vect S_i \times \vect S_j), 
\label{eq:HDMI}
\end{equation}
where $\vect D_{ij}$ is called the DM vector.  
The competition between the symmetric Heisenberg interaction and the DMI
can induce non-collinear spin configurations, such as spin spirals and magnetic skyrmions \cite{Bogdanov2001,Muhlbauer2009,Yu2010b,Tokura2021}. 
These nontrivial magnetic textures can induce ``real-space'' Berry curvature in the electron band structure, 
leading to the so-called topological Hall effect (THE) \cite{Taguchi2001}. 
Later in this review, these nontrivial magnetic structures will be employed to generate a novel form of ``frustration'', inducing novel states.

\begin{figure*}
\begin{center}
\includegraphics[width=1.9\columnwidth, clip]{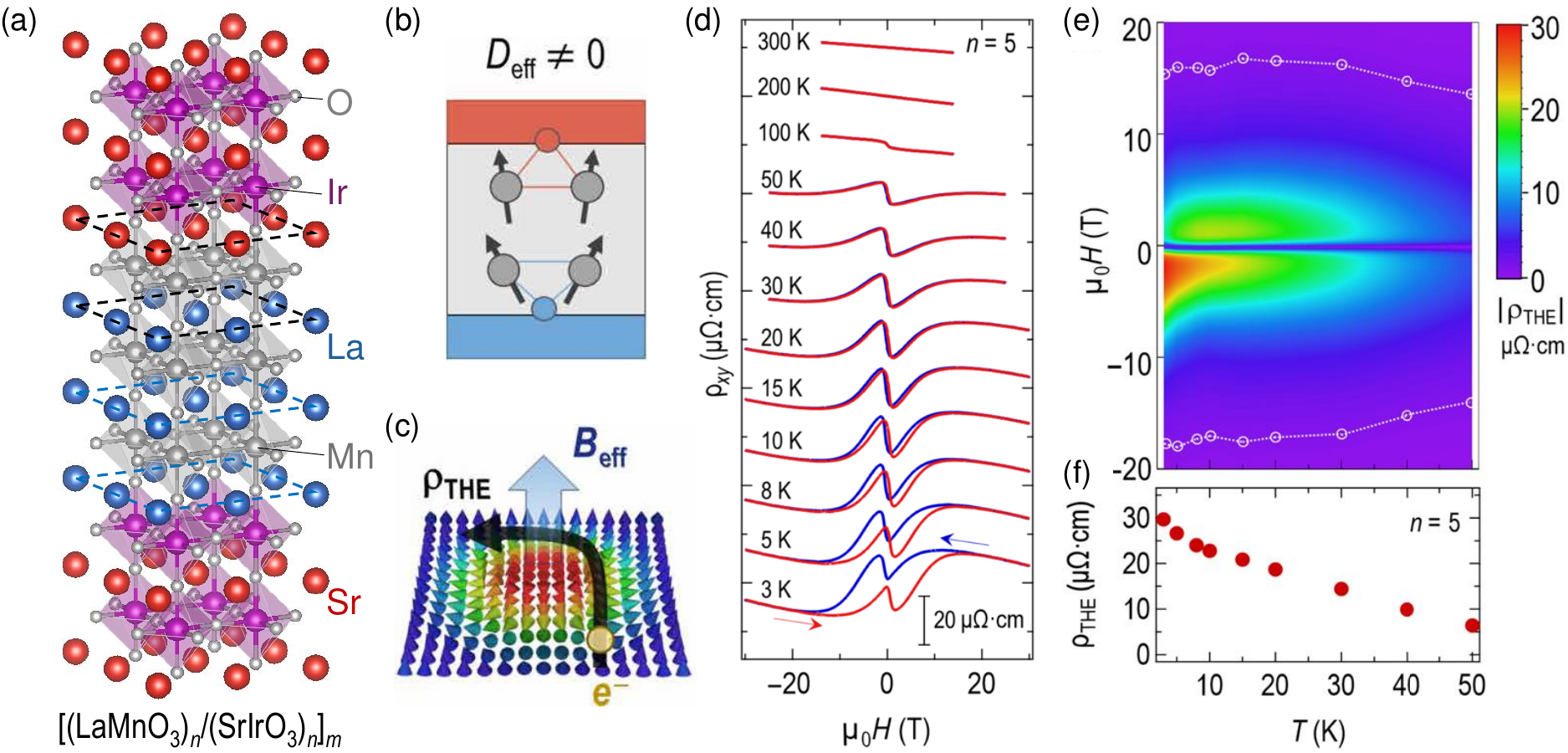}
\caption{Interfacial engineering of the DMI to control chiral magnetism in oxide superlattices.
(a) Control of the inversion symmetry in LaMnO$_3$/SrIrO$_3$ superlattices by artificially modifying the interfacial termination. 
Conventional growth of a superlattice with a BO$_2$-type layer termination creates different top and bottom interface structures, 
leading to nonzero $D_{eff}$ as shown in (a). 
(b) A schematic illustrating an effective DMI stabilized by chiral spin textures from multiple interfaces ($D_{eff}$). 
(c) A schematic of the THE ($\rho_{\rm THE}$) that results from the Berry phase accumulated by a charge carrier traversing a chiral spin texture.
The latter acts as an emergent effective magnetic field ($B_{\rm eff}$) in real space. 
(d) Temperature dependence of the Hall resistivity ($\rho_{xy}$) for the $n = 5$~u.c. superlattice varying the magnetic field. 
(e) The topological Hall resistivity ($\rho_{\rm THE}$) shown as functions of temperature and applied field. 
(f) Temperature dependence of the maximum value of $\rho_{\rm THE}$. 
Figure 1 (a) is created using VESTA software \cite{vesta}. 
Other figures adopted from Ref.~\cite{Skoropata2020}.}
\label{fig:topologicalHall}
\end{center}
\end{figure*}

{\it Topological Hall effect in TMO heterostructures}: 
By controlling the stacking sequence of constituent materials, 
the spatial inversion symmetry can be tuned, making TMO heterostructures a playground for exploring the THE 
associated with chiral magnetic textures. 
The THE in TMO heterostructures has been reported for SrRuO$_3$/SrIrO$_3$ bilayers \cite{Matsuno2016} 
(also see Refs.~\cite{Kan2018,Kimbell2020,Roy2023} for alternative interpretations), and 
(La,Sr)MnO$_3$/SrIrO$_3$ \cite{Li2019} and LaMnO$_3$/SrIrO$_3$ heterostructures \cite{Skoropata2020}.
Figure \ref{fig:topologicalHall} summarizes the experimental results for LaMnO$_3$/SrIrO$_3$ \cite{Skoropata2020}. 
As shown in Fig.~\ref{fig:topologicalHall}~(a), such a heterostructure naturally breaks inversion symmetry, 
leading to a net DMI, $D_{eff} \ne 0$, as illustrated in Fig.~\ref{fig:topologicalHall}~(b). 
The resulting chiral spin texture, skyrmions, induces a real-space Berry curvature, acting as an effective magnetic field $B_{eff}$​, 
and induces topological Hall resistivity $\rho_{THE}$​, see Fig.~\ref{fig:topologicalHall}~(c). 
In the actual experiment, $\rho_{THE}$​ appears in the hysteresis loops of the Hall resistivity $\rho_{xy}$​ as shown 
in Fig.~\ref{fig:topologicalHall}~(d). 
In panels (e,f), $\rho_{THE}$​ is shown to increase with decreasing temperature. 
In addition to the THE, the magnetic proximity effect in (La,Sr)MnO$_3$/SrIrO$_3$ heterostructures can induce spin polarization in the SrIrO$_3$ region, 
leading to a large anomalous Hall effect due to the significant Berry curvature in the presence of SOC \cite{Yoo2021}.

{\it Semi-Dirac bands in TMO heterostructures}: 
In addition to breaking inversion symmetry at interfaces, lattice symmetry in oxide heterostructures can be controlled by introducing a strain field from a substrate. 
For itinerant electron systems, the band structure or topology and lattice symmetry are tightly connected.
One important consequence of such an interplay when in the presence of SOC is the emergence of nontrivial band topology, 
leading to topological insulators \cite{Murakami2003,Kane2005,Bernevig2006a,Bernevig2006b,Fu2007,Konig2007,Hsieh2009,Hasan2010}. 
Possible topological insulators in TMO heterostructures have been explored \cite{Xiao2011,Ruegg2011,Yang2011,Okamoto2014,Okamoto2018,Hirai2015}.
In a closely related subject, band degeneracy protected by spatial symmetry has been a fundamental topic of interest for understanding 
the physics of various topological semimetals, such as Dirac, Weyl, and nodal loop semimetals \cite{Herring1937,Wan2011,Burkov2011,Young2012,Young2015,Burkov2016,Bzdusek2016,Chiu2016,Armitage2018,Burkov2018}. 
Protected band degeneracy is also useful for efficient electronics utilizing high carrier mobility associated with 
$\vect k$-linear band dispersions.

For the TMO heterostructures, pioneering work predicted semi-Dirac dispersions in TiO$_2$/VO$_2$ nanostructures \cite{Pardo2009,Pardo2010}. 
Following this proposal, an experiment demonstrated \cite{Shibuya2010} that TiO$_2$/VO$_2$ superlattices undergo 
a metal-insulator transition due to a structural transition, similar to what occurs in bulk VO$_2$. 
Thus, precise control of the lattice structure is key to realizing novel band structures. 
Dirac semimetallic behavior has been observed in bulk perovskite CaIrO$_3$ with high carrier mobility ($6 \times 10^4$\,cm$^2$V$^{-1}s^{-1}$) \cite{Fujioka2019}. 
Additionally, a novel topological semimetallic state was predicted for bulk perovskite SrIrO$_3$ with an orthorhombic structure, 
where bulk Dirac nodal rings and surface zero-energy modes are protected by mirror reflection symmetry \cite{Chen2015}.

\begin{figure*}
\begin{center}
\includegraphics[width=1.9\columnwidth, clip]{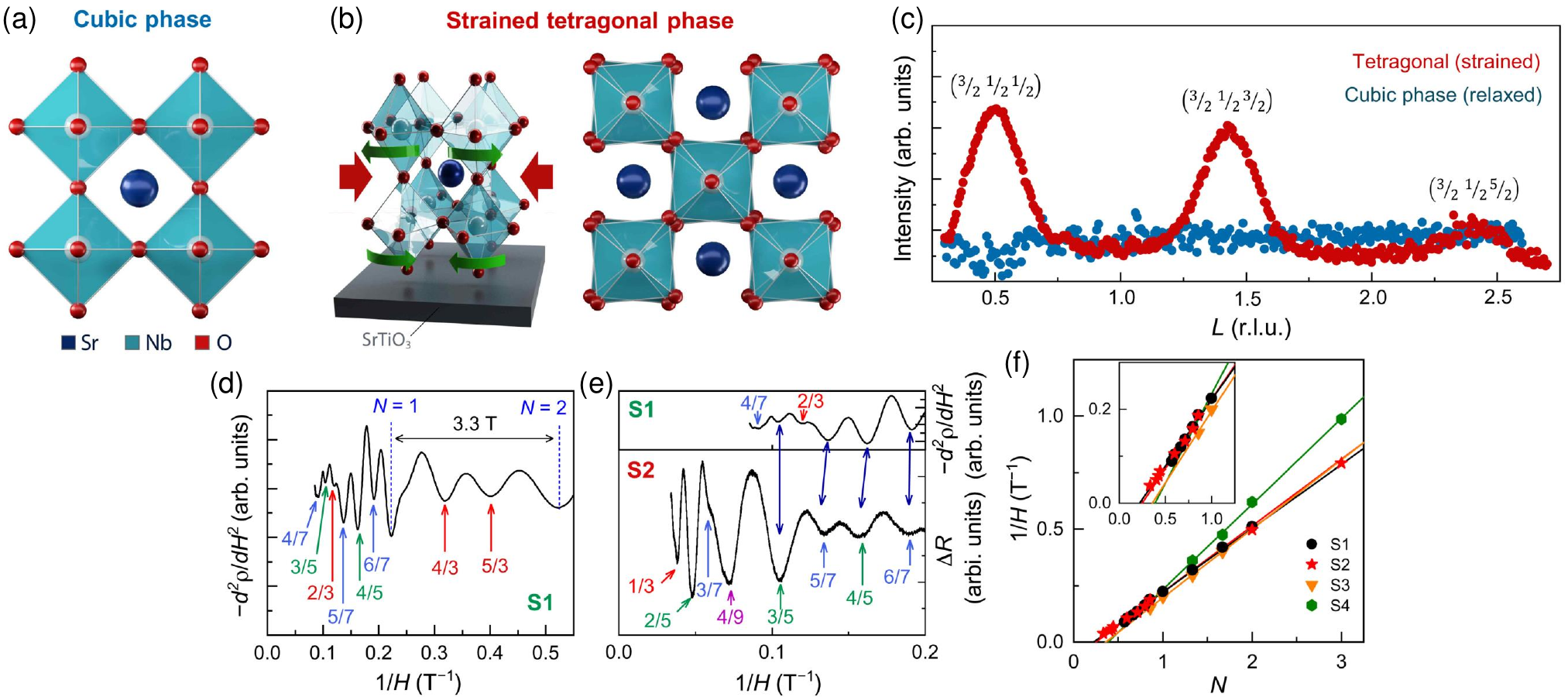}
\caption{
Strain-induced Dirac metallic state in SrNbO$_3$ thin films.
Octahedral distortion patterns for (a) cubic SrNbO$_3$ ($a^0 a^0 a^0$ in the Glazer notation) 
and (b) strained tetragonal SrNbO$_3$ ($a^0 a^0 c^-$). Epitaxial strain induces octahedral distortion. 
(c) Octahedral rotation-induced half-order superstructure diffraction peaks of 
$(\nicefrac{3}{2} \, \nicefrac{1}{2} \, \nicefrac{L}{2})$ with $L = 1, 3, 5$ for fully strained 
(red, 7.2~nm, $c^-$ rotation) and fully relaxed (blue, 130~nm, $c^0$ rotation) SrNbO$_3$ thin 
films. r.l.u., reciprocal lattice unit.
(d,e) Results of quantum oscillation measurements with different samples, S1 and S2. %in the quantum limit.
(d) shows $−d^2 \rho /dH^2 $ as a function of $1/H$ under a magnetic field of up to 14 T for S1. 
The resistivity minima are assigned as integer (fractional) Landau levels, as indicated by the arrows. 
(e) shows $\Delta R$ as a function of $1/H$ under a magnetic field of up to 30 T for S2. S1 and S2 
samples show consistent behavior. 
(e) Landau fan diagram of the Landau level index $N$ versus $1/H$ for four different samples; 
the inset shows an enlarged view of the high-field region. 
All samples show linear behavior and $1/H$ vs. $N$ lines intersect $1/H=0$ at nonzero $N$, 
indicating a nontrivial Berry phase.
The figure is adopted from Ref.~\cite{Ok2021}.}
\label{fig:semidiracexp}
\end{center}
\end{figure*}

Recently, high-quality thin films of SrNbO$_3$  were fabricated, and it was 
observed that their carrier mobility is very high ($10^3$--$10^5$\,cm$^2$V$^{-1}s^{-1}$) \cite{Ok2021}. 
As summarized in Fig.~\ref{fig:semidiracexp}, their magnetotransport results suggest 
that a Dirac semimetallic state is realized in strained SrNbO$_3$. 
This material
SrNbO$_3$ has a cubic perovskite structure when in bulk form, Fig.~\ref{fig:semidiracexp}(a), 
but a tetragonal distortion appears under compressive strain (b), 
as indicated by X-ray diffraction measurements (c). 
The tetragonal distortion is characterized by half-order 
peaks at $(\nicefrac{3}{2} \, \nicefrac{1}{2} \, \nicefrac{L}{2})$ of X-ray diffraction 
indicating a $c^-$ rotation of the NbO$_6$ octahedra. 
From magnetotransport measurements (d,e), a Landau fan diagram was constructed 
as a function of the Landau level index $N$ and $1/H$ (f). 
As shown in panel (f), the $1/H$ vs. $N$ lines intersect $1/H=0$ at nonzero $N$, 
indicating a nontrivial Berry phase originating from Dirac-type band crossings.

We also note that recent theoretical work showed that an orthorhombic distortion 
is energetically more stable than a tetragonal distortion~\cite{Rosendal2023}. 
Thus, stabilizing metastable phases during the thin-film growth process is 
crucial for uncovering novel electronic states.

%{\color{red} 
{\it Lattice distortion and magnetism}: 
As discussed in the previous paragraphs, lattice distortion is quite common in TMOs and can be controlled by heteroengineering. 
In many cases, TM-O-TM bonds are distorted, and the inversion center, originally located at O site in the middle, is lost. 
When additional N{\'e}el-type antiferromagnetic (AFM) ordering is present on TM sites, 
momentum-dependent spin splitting appears in the electron band structure.  
This novel possibility was first pointed out for TMOs \cite{Noda2016,Okugawa2018}, 
and later many materials have been identified to host such spin-split bands. 
This new class of magnetism is now called ``altermagnetism.'' 
Since many TMOs are magnetic, heteroengineering could play an important role in this new research area. 
We will discuss this possibility later in this article. 
%}

The purpose of this article is to present our theoretical work on TMO heterostructures, 
focusing on the novel phenomena induced by controlling the lattice symmetry and geometry. 
Some of the efforts reported here aim  to understand the physics behind the above experimental results. 
We also discuss future perspectives stimulated by the current topics and related research areas. 
The rest of this article is organized as follows:  
In Sec.~\ref{sec:results}, we focus on superlattices of magnetic oxides. 
First, we discuss THE in manganite/iridate heterostructures~\cite{Mohanta2019} 
relevant to the above experimental results.  We then discuss further complexities 
in such magnetic heterostructures, which could induce novel magnetic and electronic phase behaviors. 
In Sec.~\ref{sec:semidirac}, we switch to strained oxides to discuss novel electronic 
behavior controlled by the lattice symmetry, such as semi-Dirac and Weyl semimetals \cite{Mohanta2021}. 
The possibility of %{\color{red} 
metallic altermagnetism in such strained TMOs is also discussed here. 
In Sec.~\ref{sec:tsc}, we consider trilayers of oxides grown along the [111] axis 
as an example of heteroengineering of lattice geometry in TMOs 
and discuss the possibility of higher-order topological superconductivity.

\section{Novel electronic phases in superlattices of magnetic oxides}
%\section{Results}
\label{sec:results}

\subsection{Skyrmions and topological Hall effects in manganite/iridate heterostructures}
\label{subsec:skyrmion}

\begin{figure*}
\begin{center}
\includegraphics[width=1.9\columnwidth, clip]{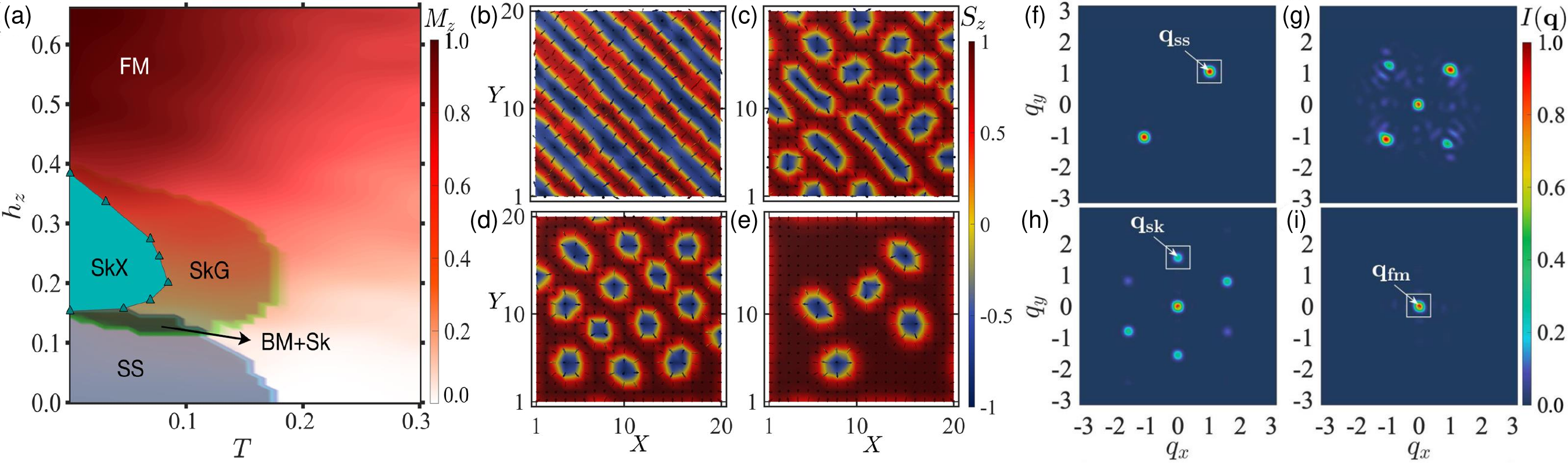}
\caption{(a) Temperature ($T$) vs. magnetic field ($h_z$) phase diagram, 
revealing spin-spiral (SS), mixed bimeron + skyrmion (BM + Sk), skyrmion crystal (SkX), skyrmion gas (SkG), and ferromagnetic (FM) phases. 
The color bar represents the normalized average magnetization, perpendicular to the interface plane. 
Typical real-space spin configurations obtained during the Monte Carlo time evolution on a $20 \times 20$ cluster with open boundary conditions
at different magnetic fields and temperatures given by (b) $h_z = 0$, $T = 0.001$, (c) $h_z = 0.2$, $T = 0.001$, (d) $h_z = 0.27$, $T = 0.001$, 
and (e) $h_z = 0.27$, $T = 0.15$, 
corresponding to the SS, BM + Sk, SkX, and SkG phases, respectively. 
The arrows denote the in-plane components $(S_{xi}, S_{yi})$ while the color bar represents the normalized perpendicular component of the magnetization $S_{zi}$.
Monte Carlo averaged intensity profile of the Bragg intensity $I(\vect q)$ for magnetic field values (f) $h_z = 0$ (SS), (g) $h_z = 0.2$ (BM), (h) $h_z = 0.27$ (SkX), and (i) $h_z = 0.37$ (FM), characterized by $\vect q_{ss}$, different $\vect q_{ss}$'s and $\vect q_{sk}$, $\vect q_{sk}$, and $q_{fm} = \vect 0$, respectively. 
In these calculations, the strength of DMI, the hopping amplitude, and the easy-plane anisotropy are fixed to $D=0.1$, $t_0 = 1$, and $A = 0.05$, respectively.
The figure is reproduced from Ref.~\cite{Mohanta2019}. 
}
\label{fig:skyrmionphasediagram}
\end{center}
\end{figure*}

In this subsection, we theoretically investigate the formation of the skyrmion crystal (SkX) phase at the (La,Sr)MnO$_3$/SrIrO$_3$ interface, 
as well as the influence of noncollinear spin textures on the transverse Hall conductance. 
Note that(La,Sr)MnO$_3$ is widely used by experimentalists in the field of oxide superlattices.
For these purposes, we analyze a spin-fermion model, where itinerant electrons and localized spins are coupled via Hund interaction, 
using a Monte Carlo (MC) simulation with the Metropolis algorithm, as detailed in Ref.~\cite{Mohanta2019}. 
We confirmed the presence of a SkX phase with N{\'e}el-type skyrmions within a range of magnetic fields, at low temperatures. 
Thermal fluctuations give rise to a related phase with spatially disordered nucleated skyrmions, known as a skyrmion gas (SkG), 
which prevails outside the SkX phase at higher temperatures and acts as a precursor to the SkX phase upon cooling. 
Additionally, a metastable phase with mixed bimerons and skyrmions (BM + Sk) appears at finite temperatures 
between the  spin-spiral (SS) and SkX phases.
For more detail and references about bimerons see Ref.~\cite{Mohanta2019}.

Specifically, we consider a square lattice which hosts the essential features of (La,Sr)MnO$_3$ at the two-dimensional interface with SrIrO$_3$. 
Namely, it contains
the intrinsic magnetism of the manganite layer, supplemented by the DMI SOC induced by the proximity with the iridate layer. 
To describe the hopping of electrons, we use a two-orbital double-exchange Hamiltonian at infinite Hund’s coupling. 
As explained before, the DM interaction arises from the influence of the iridate layer. 
Then, the resulting spin-fermionic Hamiltonian for the (La,Sr)MnO$_3$/SrIrO$_3$ interface is given by
\begin{eqnarray}
{\cal H} \!\!&=&\!\! - \!\! \sum_{\langle ij \rangle, \alpha, \beta} \biggl( t_{ij}^{\alpha\beta} \Omega_{ij} c_{i \alpha}^\dag  c_{j \beta} + {\rm H.c.}\biggr) - \mu \sum_{i, \alpha} c_{i\alpha}^\dag c_{i\alpha} \nonumber \\
&& \!\! -h_z \sum_i S_i^z + A \sum_i |S_i^z|^2  + {\cal H}_{\rm DMI}. 
\label{eq:HSF}
\end{eqnarray}
Here, $c_{i \alpha}^\dag$ ($c_{i \alpha}$) is the electron creation (annihilation) operator at position $\vect r_i$ and orbital $\alpha$, and  
$t_{ij}^{\alpha\beta}$ is the hopping amplitude between orbital $\alpha$ at $\vect r_i$ and orbital $\beta$ at $\vect r_j$.  
For $\vect r_i$ and $\vect r_j$ along the $x$ direction, $t_{ij}^{\alpha\beta}$ takes the following form: 
$t_{ij}^{aa} = 3 t_{ij}^{bb} = 3 t_{ij}^{bb} =−\sqrt{3} t_{ij}^{ab (ba)} = 3 t_0/4$, with 
$a$ and $b$ representing the Mn $e_g$ orbitals, $d_{x^2-y^2}$ and $d_{3z^2-r^2}$, respectively \cite{Slater1954}. 
$t_0$ is the hopping amplitude between $d_{3z^2-r^2}$ orbitals along the $z$ direction with an oxygen $p_z$ orbital in the middle. 
$t_0$ will be taken as the unit of energy in the rest of this subsection. 
Reflecting the sign change of $d_{x^2-y^2}$ between the $x$ and $y$ directions, 
$t_{ij}^{ab (ba)}$ along the $y$ direction has the opposite sign. 
$\Omega_{ij} = \cos(\theta_i/2) \cos(\theta_j/2)+ {\rm e}^{−{\rm i} (\phi_i - \phi_j)} \sin (\theta_i/2) \sin (\theta_j/2)$ 
represents the renormalization of hopping due to the infinite Hund’s coupling between itinerant electrons and localized spins $\vect S_i$, 
which are treated as classical vectors using the canonical polar and azimuthal angles, $\theta$ and $\phi$, respectively,  
as $\vect S_i = S (\cos \theta_i \sin \phi_i, \sin \theta_i \sin \phi_i, \cos \theta_i)$ with the amplitude $S=3/2$. 
The coupling $A$ represents the easy-plane anisotropy, originating from interfacial strain and Rashba SOC, 
which is not explicitly included in the Hamiltonian above. 
$h_z$ is the external magnetic field, applied perpendicular to the interface plane. 
${\cal H}_{\rm DMI}$ is given by Eq.~(\ref{eq:HDMI}). 
For the current geometry, reflecting the mirror symmetry with respect to nearest-neighbor (NN) bonds, i.e., ${\cal M}_{x,y}$, 
the DM vector is given by 
$\vect D_{ij} = D \hat z \times \vect r_{ij} /|\vect r_{ij}|$ with $D$ being the DMI strength and $\vect r_{ij} = \vect r_i - \vect r_j$.

We analyze our theoretical model via classical MC simulations (for details, see Ref.~\cite{Mohanta2019}). 
In addition to the spin configuration, we compute the skyrmion number $N_{sk}$, spin structure factor $S_{\vect q}$, 
as well as Hall conductivity $\sigma_{xy}$. 
The skyrmion number is given by 
\begin{equation}
N_{sk} = \frac{1}{4 \pi} \int dx dy \, \vect S \cdot \bigg( \frac{\partial \vect S}{\partial x} \times \frac{\partial \vect S}{\partial y} \bigg). 
\label{eq:Nsk}
\end{equation}
The spin structure function is given by 
\begin{equation}
S_{\vect q} = \frac{1}{N} \sum_{ij}^{|\vect r_{ij}|< \delta} \langle \vect S_i \cdot \vect S_j \rangle  \, 
{\rm e}^{-{\rm i} \vect q \cdot \vect r_{ij}}, 
\label{eq:Sq}
\end{equation}
where $\delta$ is the radius of a circle around site $i$ within which all sites are considered 
to calculate the real-space correlation function
$\langle \vect S_i \cdot \vect S_j \rangle$, 
and $N=N_x \times N_y$ is the total number of lattice sites. We use the radius $\delta$ up to $N_x/2$. 
The Hall conductivity is given by 
\begin{equation}
\sigma_{xy} = \frac{e^2}{h} \frac{2 \pi}{N} \sum_{\vect k, m} \Omega_m^z (\vect k) f(\varepsilon_{m \vect k}), 
\label{eq:sigmaxy}
\end{equation}
where 
$\Omega_m^\alpha (\vect k)$ is the $\alpha$ component of the Berry curvature vector of the $m$-th band, which is given by 
\begin{eqnarray}
\Omega_m^\alpha (\vect k) = - \!\!\!  \sum_{n}^{\varepsilon_{n \vect k} \ne \varepsilon_{m \vect k}} \!\!\! \varepsilon_{\alpha \beta \gamma}
{\rm Im} \frac{
\langle \psi_{m \vect k} | v_{\vect k}^\beta| \psi_{n \vect k} \rangle \langle \psi_{n \vect k} | v_{\vect k}^\gamma | \psi_{m \vect k} \rangle 
}{(\varepsilon_{m \vect k} - \varepsilon_{n \vect k})^2}, 
\label{eq:BCV}
\end{eqnarray}
with 
$\varepsilon_{\alpha \beta \gamma}$ being the Levi-Civita antisymmetric tensor, 
$\{\alpha, \beta, \gamma\} \in \{x,y,z\}$, and $v_{\vect k}^\alpha$ being the velocity operator \cite{Xiao2010}. 

A typical phase diagram is shown in Figure~\ref{fig:skyrmionphasediagram}~(a), 
revealing five distinct phases: SS (spin spiral), SkX (skyrmion crystal), BM + Sk (bimeron + skyrmion), SkG (skyrmion gas), and FM (ferromagnetic), as discussed before.
These phases exhibit characteristic spin configurations, with snapshots shown in panel (b) for SS, (c, d) for SkX, and (e) for SkG. 
The spin structure factors are depicted in panel (f) for SS, (g) for BM, (h) for SkX, and (i) for FM.
The SS and the SkX phases show, respectively, the single-$\vect q$ (b)
and triple-$\vect q$ (d) structures of the spin configuration. 
Plot (c), for the BM+ Sk phase, displays a double-$\vect q$ spin configuration, which is absent at $T = 0$. 
The FM phase is identified by both the Bragg intensity $I(\vect q_{fm})$ and the average out-of-plane magnetization $M_z$. 
As shown in (f), skymions do not form a regular lattice structure in the SkG phase, thus this phase is not characterized by $S_{\vect q}$. 
Nevertheless, as shown next, this phase has finite $N_{sk}$, which induces a nonzero $\sigma_{xy}$. 

\begin{figure}
\begin{center}
\includegraphics[width=0.85\columnwidth, clip]{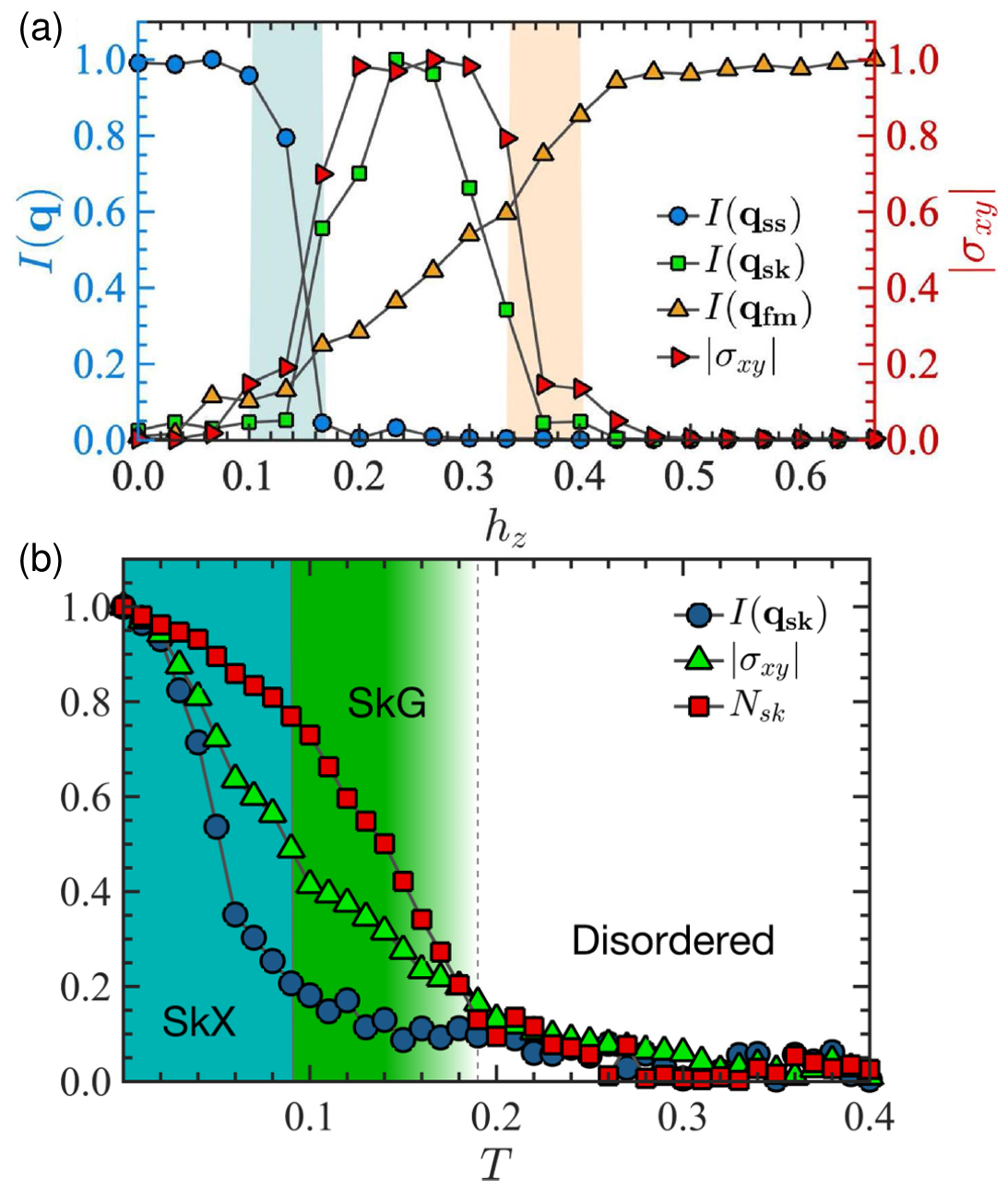}
\caption{Evolution of the topological Hall effect characterized by a finite $\sigma_{xy}$ induced by the skymion formation as functions of (a) $h_z$ at $T=0.01$ and (b) $T$ at $h_z=0.27$. 
In the SkG phase, the Bragg intensity $I(\vect q_{sk})$ is reduced, but the skyrmion number $N_{sk}$ remains finite, leading to a nonzero Hall conductivity $\sigma_{xy}$.
This figure is reproduced from Ref.~\cite{Mohanta2019}. }
\label{fig:skyrmiontransport}
\end{center}
\end{figure}

Figure~\ref{fig:skyrmiontransport} summarizes the evolution of the magnetic phases and its relation to $\sigma_{xy}$. 
At low $T$, the Bragg intensity $I(\vect q)$ at different characteristic momenta reveal a sequence of phase transitions that occur 
by changing $h_z$,  as shown in (a). 
Evidently, $\sigma_{xy}$ becomes enhanced within a broader range of $h_z$ than $I(\vect q_{sk})$, 
where $\vect q = \vect q_{sk}$ is the characteristic momentum for the SkX phase. 
%
%The SS and SkX phases can, therefore, be identified using I(qss) and I(qsk), respectively, 
%whereas, σxy and Nsk are suitable to identify the BM + Sk and the SkG phases. 
%
Panel (b) contains plots of $\sigma_{xy}$ and the skyrmion number $N_{sk}$ as functions of $T$ at $h_z = 0.27$, 
namely in a region where the SkX phase is stable at the lowest $T$ . 
Evidently, $I(\vect q_{sk})$ drops faster with $T$ than $\sigma_{xy}$ and $N_{sk}$, indicating that the SkX phase 
exists at much lower temperatures than the SkG phase. 
The critical temperatures for the SkX and the SkG phases at $h_z = 0.27$ are roughly $T_c^{SkX } \simeq 0.09$ and $T_c^{SkG} \simeq 0.19$, respectively. 
The THE, although it is the strongest in the SkX phase, also exists at temperatures much above the SkX phase. 
There could be other contributions to the THE, and one potential origin is the skew scattering induced by the scalar spin chirality. 
The skew-scattering induced THE appears near the transition to the SkX phase and is reflected by a change in the sign of the Hall conductance.

In Ref.~\cite{Mohanta2020}, the spin dynamics is investigated using an effective model, where the ferromagnetic Heisenberg model is explicitly considered between localized spins, 
instead of a double-exchange mechanism induced by the Hund coupling between localized spins and itinerant electrons. 
It was shown that a characteristic spin excitation spectra appears depending on the magnetic phase. 
This behavior could be useful for identifying the underlying magnetic state and confirming the existence of the SkX phase through inelastic neutron scattering spectroscopy. 
However, measurements with these small samples are likely a factor 10 beyond current experimental capabilities, 
such as the Spallation Neutron Source in the United States, the Institut Laue-Langevin in France, 
the ISIS neutron source in the United Kingdom, and J-PARC in Japan. 
There are additional considerations such as neutron absorption, the background from the substrate scattering, bandwidth of the excitations, achieving appropriate coverage in reciprocal space, {\it etc.} for a successful neutron-scattering experiment. 
Nevertheless, such experiments are likely feasible at planned or under construction next generation neutron sources, 
such as the European Spallation Source or the Second Target Station at the Spallation Neutron Source, Oak Ridge National Laboratory, 
where improvements in instrumentation and neutron flux are anticipated to result in performance gains in excess of an order of magnitude. 
For further details on these future experimental aspects, please see Ref.~\cite{Mohanta2020}.

%\section{Novel physical concepts that require superlattices}

\subsection{A new kind of ``frustration'': General framework}

After addressing novel concepts involving skyrmions, now we move to the study of superlattices where
a novel frustrating mechanism could be found and produce other novel states.
It is well known that spin frustration in bulk materials can arise primarily from two reasons. One is the presence of
competing interactions. For example, Heisenberg models can have the canonical
NN interactions, with coupling $J_1$, among neighboring spins. But for geometrical
reasons and for overlaps of orbitals of all the atoms participating in the chemical formula, a realistic Hamiltonian
may also require interactions at longer
distances, such as next-NN $J_2$ or next-next-NN $J_3$, {\it etc}. Each of these couplings, if alone in the Hamiltonian,
would lead to different spins patterns. When all are in combination, it is difficult to predict
what spin arrangement  will emerge as ``compromise'' ground state from these different tendencies that cannot all be
simultaneously satisfied. Even spin liquids~\cite{elbio1-1,elbio1-2,elbio1-3,elbio1-4}, with no long-range spin order, could arise from this type of frustration.
 
A second common source of frustration is the shape of the lattice. For example, for an AFM
Ising model, it is impossible to have a triangle of spins where each of the three bonds is
antiferromagnetically ordered. The situation is more dramatic, of course, on an entire two-dimensional
triangular lattice. Once again, a compromise among the spins must be found and often only numerical techniques
are suitable to find the true ground state under these types of frustration.

\begin{figure*}
\begin{center}
\includegraphics[width=1.8\columnwidth, clip]{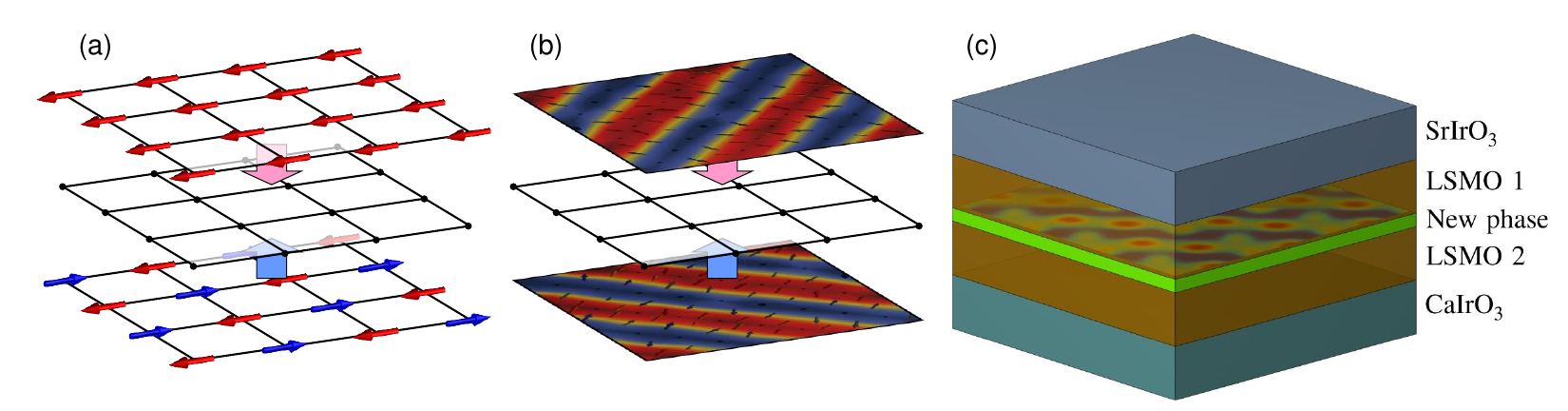}%{elbio-figure1.jpg}
\caption{
Schematic illustration of a multilayer to realize ``interfacial
phase frustration''. %from Ref.~\cite{elbio1}. 
Examples of possible realization induced by the competition between FM and AFM layers (a) and by the competition 
between spin spirals with different orientations (b). 
%Shown is 
(c) A magnetic multilayer, involving
different $5d$ compounds at top and bottom, and LSMO layers from Ref.~\cite{elbio1}. 
Our theory predicts that an
emergent novel phase will appear in the middle layer (green), between
two LSMO layers each influenced by two different $5d$ compounds, namely
with different DMI strengths or different spiral orientations. The
central ``green'' layer is the most frustrated by the competing tendencies from above
and below.
}
\label{fig:elbio-figure1}
\end{center}
\end{figure*}

Recently, a new general form of frustration has been proposed where exotic magnetic phases
can be created using ``interfacial phase frustration'' via artificially grown superlattices~\cite{elbio1}.
The primary idea is easy to explain. Consider a multilayer system where the top and bottom layers
are fixed to configurations that are {\it not} compatible with one another, such as 
having FM ordering at the top layer and AFM ordering at the bottom layer as illustrated in Fig.~\ref{fig:elbio-figure1} (a), 
and having stripes in one
direction at the top and in a perpendicular direction at the bottom as discussed in depth later as illustrated in Fig.~\ref{fig:elbio-figure1} (b). 
The latter situation could be realized in superlattices involving manganites and iridates, Fig.~\ref{fig:elbio-figure1} (c). 
All the layers in-between will need
to re-adjust their spin orientation to partially satisfy the constraints imposed at the top and bottom,
thus inducing new twisted magnetic and charge orders that, often, do {\it not} exist in real individual materials and
can only be induced in the above-mentioned frustrated superlattices. In particular, 
{\it the layer exactly in the middle} between the top and bottom fixed spin configurations is the most affected, 
and where the most exotic states are expected. 
This method provides a novel playground to realize unconventional
magnetic phases that do not exist in single crystals. Of course, while it is easy to create the ``bottom'' layer by growing particular
spin configurations on a fixed substrate, placing the floating fixed top competing layer is challenging and
hopefully experimentalists could find a way to handle this issue, motivated by the theoretical results
reviewed here. As additional motivation, we have found that sometimes in these frustration-induced novel phases the spin
chirality-driven topological Hall conductivity can be largely enhanced~\cite{elbio1-5,elbio1-6}. For details and further references 
about this issue, the readers should consult the original publication Ref.~\cite{elbio1}.

While the primary idea was described in detail in Ref.~\cite{elbio1} 
here we will only focus on one example, for simplicity.
In Fig.~\ref{fig:elbio-figure1}, a sketch of the proposed platform is shown, with two different
$5d$ compounds at top and bottom, such as SrIrO$_3$ and CaIrO$_3$ with robust DMI coupling. The ensemble also 
includes metallic lanthanum manganites La$_{1-x}$Sr$_x$MnO$_3$ (LSMO), which is a widely used 
material in the field of artificially
grown superlattices. Manganites were originally much studied because of the discovery of the colossal
magnetoresistance (CMR) where relatively very small magnetic fields can change their resistivity by many
orders of magnitude~\cite{ramesh,Tokura1994}. 
Later it was discovered that manganites are ideal materials for this type of novel frustration 
ideas because of the wide variety
of phases that they display, particularly when Sr is replaced by Ca~\cite{Schiffer1995,Tokura1996,Tokura1999,elbio1-12,elbio1-13,elbio1-14,elbio1-15,hotta2000,hotta2000b,aliaga}.
The iridate components, with Ir being in the $5d$ row of the periodic table, 
contain a robust spin-orbit coupling that
we described phenomenologically in~\cite{elbio1} via a canonical DMI of strength $D$. 
Empirically, it has been known, and described
in previous portions of this review, that the DMI in the iridates can be transmitted into LSMO~\cite{elbio1-19,Skoropata2020}. 
This is similar as when a superconductor is close to a non-superconducting 
metal: superconductivity can be induced in the first few layers of the non-superconducting metal. 
This is widely known as ``proximity effect''
and is not restricted to superconductivity. Thus, in this artificial geometry, 
the LSMO layers can acquire a nonzero DMI interaction due to the proximity to a $5d$ material,
even if Mn is a $3d$ element with weak spin-orbit coupling. In this context, 
a variety of magnetic phases, including a spiral 
or a skyrmion crystal~\cite{elbio1-7,elbio1-8,elbio1-9,elbio1-10,elbio1-11,elbio1-18} 
with various radii can be induced, as our work has shown, depending on the doping level in the LSMO 
and the DMI strength transmitted at the iridate/LSMO interfaces. 

When two different magnetic phases
are realized in the two LSMO layers next to the top and bottom iridates, due to their influence 
but with top and bottom having different strengths $D$, 
an unconventional magnetic phase will arise at the center (this center layer 
is indicated with light green in Fig.~\ref{fig:elbio-figure1}). 
Our proposal was confirmed by performing MC 
calculations using a relatively simple spin Hamiltonian for a magnetic multilayer, 
given by 
\begin{eqnarray}
{\cal H} = -J\sum_{\langle ij \rangle} \vect S_i \cdot \vect S_j + {\cal H}_{\rm DMI} - A_z \sum_i |S_i^z|^2 ,
\end{eqnarray}
which contains 
%containing
ferromagnetic Heisenberg interactions (because LSMO is assumed to be in the doping range where 
ferromagnetism dominates) of strength $J$. The Hamiltonian also contains a DMI term 
induced by proximity to the iridates of strength $D$, as already discussed. 
In addition, uniaxial anisotropies of strength $A_z$, that are often present in many superlattices due to compressive strain,
%most superlattices , 
were included in the model.
%(for details once again see Ref.~\cite{elbio1}). 

In our original publication, we considered three cases and obtained three unconventional magnetic
phases in all of them: (i) a checkerboard skyrmion crystal (CSkX) that appears
from the competition between two orthogonally-aligned spin
spirals, (ii) an incommensurate skyrmion stripe (ISkS) that arises
from the competition between a spin spiral and a triangular
skyrmion crystal, and (iii) a ferrimagnetic skyrmion crystal (FSkX)
that arises from the competition between a triangular skyrmion
crystal and a standard antiferromagnet. 
The CSkX case, which will be our focus below, is particularly interesting because it is created {\it without any
external magnetic field} and is, therefore, promising for spintronic
applications. Furthermore, we found this CSkX produces a large THE, induced by a nonzero scalar spin chirality, which is
absent in the competing spin spirals. Thus, chirality can arise from
non-chiral components.

\subsection{Specific example with orthogonal spin spirals at top and bottom}

Consider two spin spirals as the top and bottom layers of the multilayer system described above. 
The spin spiral phase, generated by a finite DMI even in the absence of magnetic fields or uniaxial anisotropy, has two degenerate
spiral solutions, aligned mutually-orthogonal to each other. 
One of these two preferred directions can be realized by strain mismatching~\cite{elbio1-16,elbio1-17}, 
but considering that they are energetically degenerate, multiple trials are required to obtain the desired  and stable configuration, 
where top and bottom layers have oppositely-oriented stripes. 
In our actual MC calculations, these two spin spirals were spontaneously generated
along the preferred directions by applying a small in-plane anisotropy (for details see Ref.~\cite{elbio1})

\begin{figure*}
\begin{center}
\includegraphics[width=1.9\columnwidth, clip]{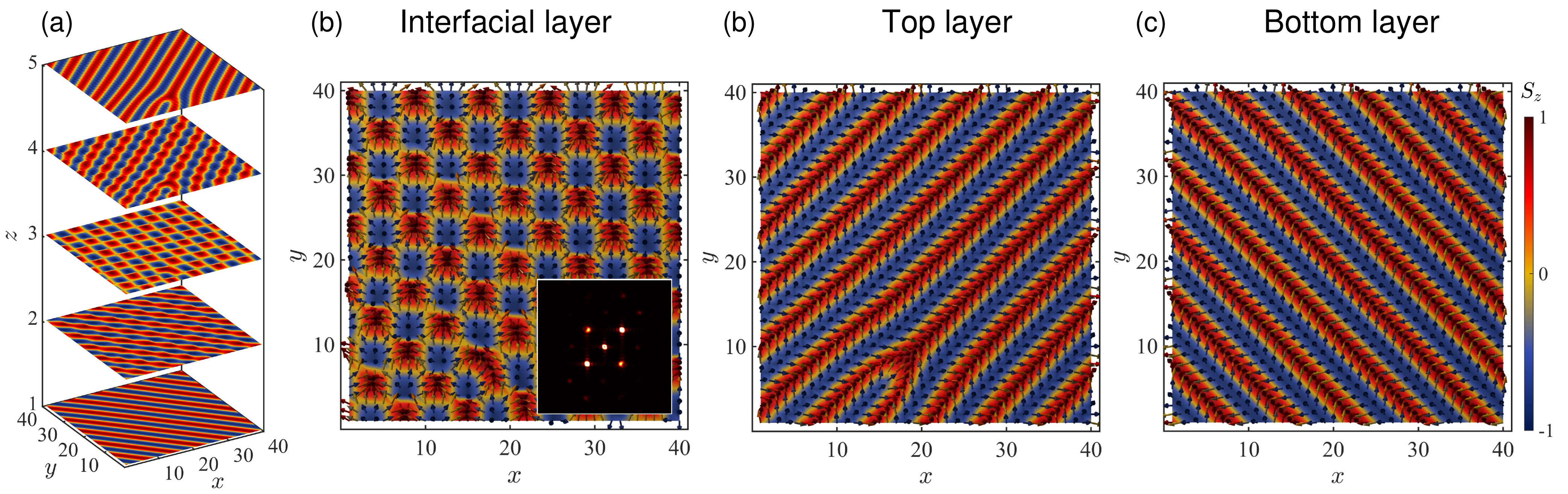}%{elbio-figure2.jpg}
\caption{
Checkerboard skyrmion crystal. (a) Plot of the $z$-component of the spin $S_z$, obtained via MC annealing, 
in a five-layer heterostructure. (b) Spin configuration of the frustration-stabilized checkerboard skyrmion 
crystal (CSkX) at the central layer ($z = 3$). This is the most  important result of this figure, because 
this phase was never stabilized in any other context than the one provided here. (c) and (d) are the canonical
spin spirals at the top layer ($z = 5$) and the bottom layer ($z = 1$), oriented along 
mutually-orthogonal diagonals. The small imperfection
in panel (c) simply shows that the configuration was MC generated, as opposed to being created by hand. 
The inset in panel (b) shows the central
layer spin structure factor $S(q)$ within the first Brillouin zone, revealing a double-$q$ magnetic order. 
The parameters used in the classical Hamiltonian MC study are $D = J$, $A = 0$, $J = 1$. For details see Ref.~\cite{elbio1}.
}
\label{fig:elbio-figure2}
\end{center}
\end{figure*}

Figure~\ref{fig:elbio-figure2} shows the spin configurations at different
layers of the considered five-layer heterostructure. The competition
between the two spin spirals, fixed at top and bottom as in panel (a), produces an unconventional
checkerboard skyrmion crystal (CSkX) at the center. This is the most surprising result of the study. 
The spin structure factor is shown
in the inset of panel (b) of Fig.~\ref{fig:elbio-figure2} and is compatible with 
the real-space main figure in panel (b) of the same figure,
where an array of {\it squared} skyrmions is shown, as opposed to the conventional circular skyrmions
forming a triangular lattice when in the absence of the new frustration effects discussed here in previous pages. 
To our best knowledge, this configuration is new and was never seen before in the context of skyrmions, highlighting the
novelty of the concepts discussed here: {\it new states can be created via artificially grown superlattices that do not have
an analog in single crystals or in films}.

We finish this subsection with technical details about the
MC simulations. We performed an ``annealing'' slow reduction in temperature from
high temperature, to obtain the ground state spin
configuration in the discussed magnetic multilayer. It is crucial to use annealing, otherwise the system 
can easily get trapped
in metastable configurations due to the complexity of the magnetic orders in the various layers. 
We consider periodic boundary conditions in plane, open boundaries in the $z$-direction, 
and use the Metropolis spin update procedure.
Specifically, during annealing we started at a high-temperature $T=5J$, where $J$ is the Heisenberg
coupling, and very slowly
reduced the temperature down to $T=0.001 J$. At each temperature, a huge
number of MC spin updates, typically of the order of $10^{11}$, were performed to
avoid being trapped in metastable states.
The ground state magnetic configurations at the top two and the bottom two
layers in our five-layer structure were first stabilized by performing MC
annealing, without any communication between the top two and bottom two
layers. Next, we switched on the interlayer coupling between all the layers and
perform MC annealing in the entire multilayer system. During this process, we started
from a completely random spin configuration at the middle layer, thus the results shown
in Fig.~\ref{fig:elbio-figure2}~(b) are totally unbiased and {\it emerge} from the 
new form of frustration described here.

%----------------------------------------------

\subsection{Prediction of novel phases in narrow-bandwidth superlattices of manganites}

Because of the simultaneous participation of several degrees of freedom in TMOs, such as 
spin, charge, orbital, and lattice, it is expected that artificial multilayer structures made of
these materials will exhibit {\it much richer} physics than in conventional semiconductor heterostructures.

\begin{figure}
\begin{center}
\includegraphics[width=1\columnwidth, clip]{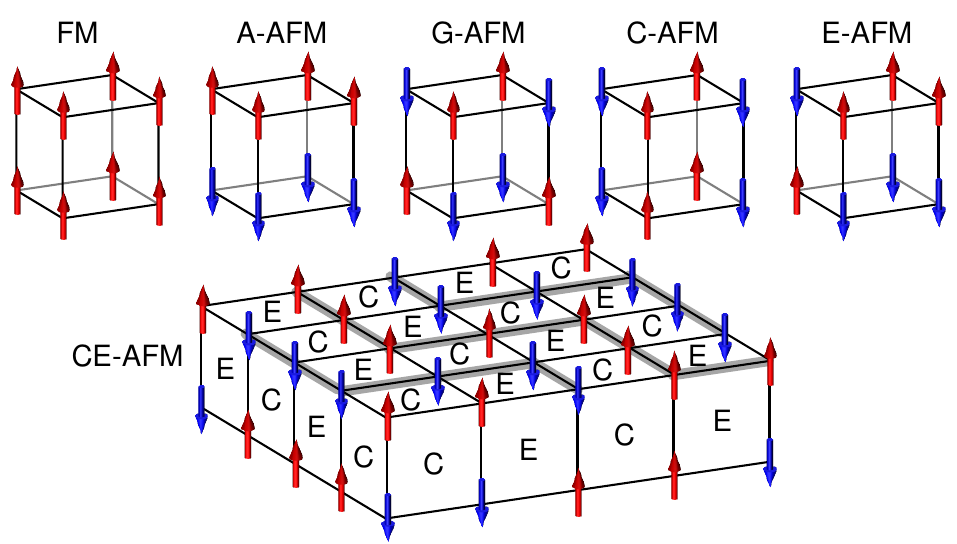}
\caption{
Schematics of magnetic ordering in perovskite manganites. 
Top row: Ferromagnet (FM), A-type AFM (A-AFM), G-type AFM (G-AFM), C-type AFM (C-AFM), and E-type AFM (E-AFM).
E-AFM pattern alone cannot fill three-dimensional lattice. 
The combination of C-AFM and E-AFM, i.e., CE-AFM, is often observed in manganites. 
}
\label{fig:magneticordering}
\end{center}
\end{figure}

Among several efforts in the recent past, there was considerable interest
in the analysis of $R$MnO$_3$/$A$MnO$_3$ ($R$MO/$A$MO) heterostructures,
where $R$=La, Pr, ... is a trivalent rare earth and $A$=Sr, Ca, ... is a divalent alkaline element. For more details
see the following Section. 
Bulk manganites exhibit a variety of magnetic ordering as schematically shown in Fig.~\ref{fig:magneticordering}.
At low temperatures, the bulk $R$MO is in a so-called A-type AFM 
state, A-AFM, involving FM layers one over the other with spins pointing in opposite directions.
Thus, while in plane they are FM, between planes they are AFM. This A-AFM state is an insulator. Meanwhile
the bulk $A$MO is in a G-type AFM state, with the canonical staggered AFM order in all three directions, 
that is also an insulator. Upon doping, experimentally it is known that the alloy $R_{1-x}$$A_x$MnO$_3$ 
does exhibit a wide variety of states depending on the doping concentration $x$, which controls 
the electronic charge density in the alloy (see Refs.~\cite{elbio1-12,elbio1-13,elbio1-14,elbio1-15,Maezono1998a,Maezono1998b,Okamoto2000} and references therein). 
Refs.~\cite{elbio1-12,elbio1-13,elbio1-14,elbio1-15} focused on Jahn-Teller type electron-lattice coupling, while 
Refs.~\cite{Maezono1998a,Maezono1998b,Okamoto2000} focused on electron-electron interactions to induce various spin and orbital ordering. 

However, the heterostructure $R$MO/$A$MO could potentially {\it behave differently} from 
its parent bulk compounds, even when they are doped. For instance,
the transfer of charge through the interface caused by the different Fermi energies
of the superlattice components  (different work functions~\cite{kancharla}), 
and concomitant different electronic
density concentrations, surely will cause a distribution of charge that is not homogeneous
along the growth direction. Hence, several states may exist in different regions of the heterostructure. As shown below, some
of those states may not even have an analog in crystals of doped manganites, which is surprising.

In previous publications, 
we used the two-orbital double-exchange model for manganites, which has been successfully applied
to the study of bulk Mn oxides for decades, to the analysis of $R$MO/$A$MO heterostructures. 
In addition to spin-dependent hopping influenced by the direction of local magnetic moments, as described by the first term in Eq.~(\ref{eq:HSF}), 
the theoretical model includes the AFM superexchange interaction between neighboring spins as described by $J^{AF} \sum_{\langle ij \rangle} \vect S_i \cdot \vect S_j$ 
and Jahn-Teller coupling (see Ref.~\cite{elbio1-14} and references therein). 
These heterostructures are assumed to be grown along the [001] direction, namely the $z$-axis. We here also
assume that the length of the heterostructure is long enough that the states at the two ends 
of the heterostructure resemble closely
their bulk counterparts, i.e., an A-AFM state at the $R$MO side and a G-AFM state at the $A$MO side. 
This assumption allowed us to simplify the analysis and focus on what state is stabilized at the only interface present in the heterostructure. 

One of the main novel results of our study is the observation of states close to the
interface that {\it do not have} an analog in experimentally known
bulk doped phase diagrams~\cite{canted-CE} 
(also, see Fig.~\ref{fig:magneticordering}). 
These states arise as interpolations between, for example, the A-AFM and CE states 
that dominate in the
bulk and interfaces, respectively. The CE state was introduced long ago by the late J. Goodenough \cite{Goodenough1995} 
and contrary to some misconceptions ``CE''
does not mean correlated electrons (see Ref.~\cite{elbio1-14} and references therein). 
It is just a notation to label various AFM states that used the first letters of the alphabet. 
CE, in particular, refers to a quite exotic state involving zigzag chains, one next
to the other. Each zigzag chain has all its spins oriented along the same direction, so each zigzag chain is FM. 
But the next zigzag chain points in the opposite direction. Thus, overall this is a complex form of AFM order, 
regarding the spin. Also each
zigzag chain has orbital order and its charge is not distributed uniformly. For details, 
we refer the reader to Ref.~\cite{elbio1-14} and references therein. Thus,
this CE exotic state is not only AFM, but it is also charge and orbitally ordered.

One of the biggest surprises of our theoretical study of heterostructures involving Ca replacing La, 
contrary to the far more common Sr replacing La,
was the observation of ``canting'' of the spins in the CE zigzag chains creating a never observed before, 
neither theoretically nor experimentally, {\it canted CE state}. In this respect,
this analysis of Ca-based manganite heterostructures lead to similar conclusions as in the previous section involving heterostructures that
produced exotic skyrmions: because of the new third form of frustration described before, the superlattices can stabilize novel
states that have never been seen before in individual crystals.

In our calculations reviewed here, the charge transfer through the interface is taken into 
account via the self-consistent solution of the charge density considering the long-range 
electrostatic Coulomb potential. For the heterostructure discussed here, 
the charge always transfers from the A-AFM side (with $x=0$) to the G-AFM side (with $x=1$), as common sense indicates. 
To properly describe the charge transfer, the long-range Coulomb interaction among the mobile
$e_g$ electrons of the manganites, as well as with the positively charged ionic background, must be included into the Hamiltonian 
at least at the mean-field level. Details can be found in Ref.~\cite{canted-CE} but essentially we solve numerically the Poisson equation
in a self-consistent manner to find the charge at each layer. This is a tedious and CPU-time consuming effort, yet conceptually
it is straightforward.
Similar theoretical phase diagrams were reported in Refs.~\cite{Maezono1998a,Maezono1998b,Okamoto2000} focusing on strong correlation effects. 

\begin{figure}
\begin{center}
\includegraphics[width=1\columnwidth, clip]{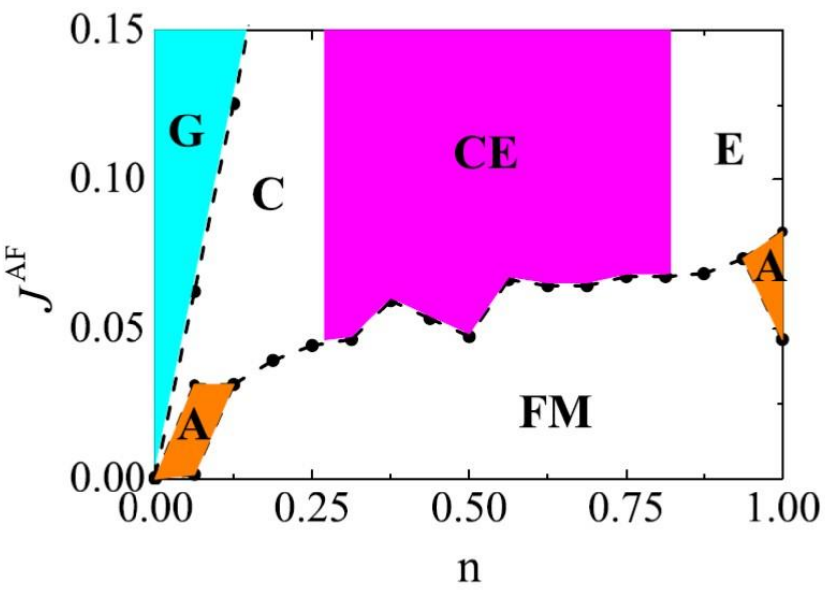}
\caption{Phase diagram of the two-orbital model solved numerically on a finite cluster
as a function of electron density $n$ and the NN AFM superexchange interaction $J^{AF}$. 
% at various densities $n$. 
The colored phases are the important ones
for our discussion, namely the A-AFM phase at $n=1$ ($x=0$), the CE phase in a vast doping region near $n=0.5$, and
the G-AFM phase close to $n=0$.
Details can be found in the original Ref.~\cite{canted-CE}. 
}
\label{fig:newfrustration1}
\end{center}
\end{figure}

To analyze the results of the heterostructures, it is important to remind the readers of the
phase diagram of the {\it bulk} two-orbital model, which is shown in Fig.~\ref{fig:newfrustration1}, using special sets of parameters
as illustration. This phase diagram is a theoretical one, arising from computational studies, but resembles the
experimental one (see for example Refs.~\cite{Tokura1994,Schiffer1995,Tokura1996,Tokura1999}). 
Once again, for fine details we refer the reader to Ref.~\cite{canted-CE} but sufficient is to notice
the vast number of phases obtained varying the electronic density. This includes
A-AFM, G-AFM, C-AFM, E-AFM, CE-AFM, and FM states. The results were numerically obtained on a finite cluster.
Note that there could exist even more exotic states in the phase diagram, such as the
C$_x$E$_{1-x}$ state (see Ref.~\cite{elbio1-14} and references therein). However, for simplicity our interest will be 
mainly on the A-AFM state stabilized at electronic density $n=1$, the CE and A-AFM states at $n=0.5$,
and the G-AFM state at $n=0$ (where $n=1-x$). We point out that the bulk phase diagram found theoretically
consists of G-AFM, C-AFM, CE, FM, and A-AFM phases consecutively from low to high electronic densities, correctly
resembling the phase diagram of real narrow-to-intermediate bandwidth bulk manganites found experimentally.

\begin{figure}
\begin{center}
\includegraphics[width=0.9\columnwidth, clip]{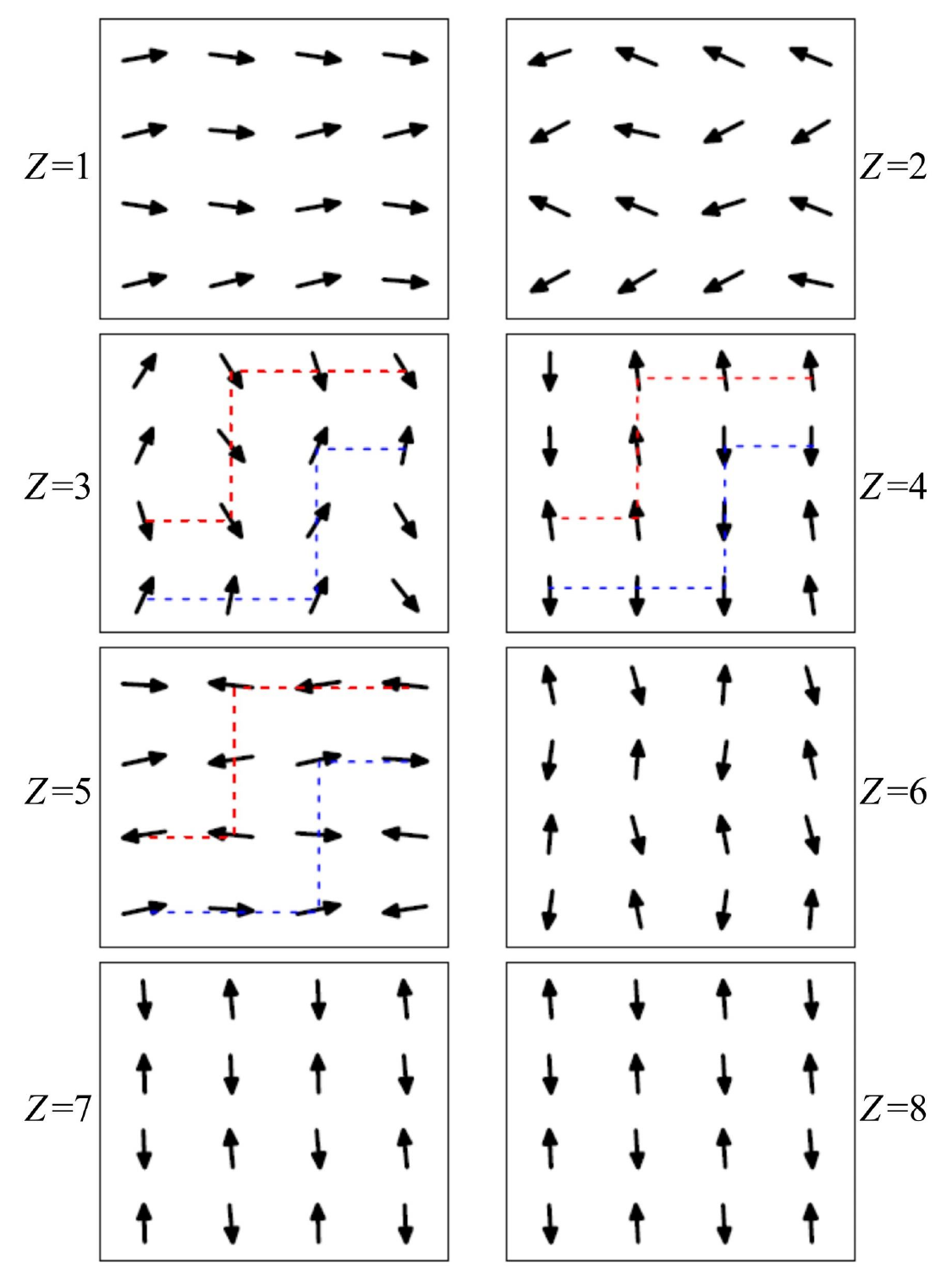}
\caption{Monte Carlo optimized classical spin configuration of a 4$\times$4$\times$8 cluster using as model
the double-exchange, characteristic of all manganites, in combination with classical Jahn Teller distortions,
which are prominent for Ca-based manganites.
Figure reproduced from Ref.~\cite{canted-CE} where more details, such as couplings used, can be found. The label $Z$ is for the 8 layers
studied simulating the direction of superlattice growth. For $Z=1,2$, the spins are very close to an A-AFM state,
as it occurs at zero hole doping in i.e. LaMnO$_3$. For $Z=6-8$, the spins are very close to a G-AFM state
as it occurs at full hole doping in i.e. CaMnO$_3$. The electronic density at layers $Z=3-5$ smoothly evolves from those two limits.
In particular, $Z=4$ corresponds to La$_{0.5}$Ca$_{0.5}$MnO$_3$ where the CE state dominates, as shown by the blue and red dotted lines guiding the eye indicating 
the zigzag chains characteristic of this state. The novelty of this effort is located at both layers $Z=3,5$, but is the
most prominent at $Z=3$. In this layer, there are clear zigzag lines resembling the CE state, 
due to the influence of layer $Z=4$. However, the $Z=3$ state has a net nonzero
magnetization along the horizontal axis, because of the influence of layers $Z=1,2$, namely the A-AFM state. The $Z=3$ result is a never-seen-before
``canted'' CE state which can only be realized in a superlattice geometry, and has no analog in bulk manganites.
}
\label{fig:newfrustration2}
\end{center}
\end{figure}

The primary ``punchline'' of our publication Ref.~\cite{canted-CE} are the states that we found numerically in our
simulated interface between the A-AFM and G-AFM states at both extremes. For the system that we solved
numerically via the self-consistent Poisson equation, the spin configurations for each of the 8 layers in the
system are shown in Fig.~\ref{fig:newfrustration2}. Consider first the layers $Z$=1 and 2. They represent the A-AFM portion of the ensemble
with $n$ close to 1. Clearly, the spins at these first two layers are primarily FM for each layer and opposite to each
other with respect to one another, as in the A-AFM arrangement, although $Z$=2 already starts showing some slight non-collinearity.
At the other extreme, consider $Z$=6,7,8. They are clearly antiferromagnetically staggered, although again $Z$=6 has
slight deviations from perfect collinearity.
Here, the total number of layers has been chosen to be 8 to investigate the effect of new frustration. 
As discussed below, this appears in a relatively narrow region, whose thickness is 3--4 layers. 
In principle, any thickness should display qualitatively similar results as long as the total thickness is much larger than 3--4. 

{\it The most interesting results are in layers $Z=3,4,5$}. First of all, the red and blue dashed lines help  in guiding
the eye for the readers to easily recognize the zigzag chains characteristic of the CE phase of bulk manganites at $n=0.5$,
even though neither the nominally $n=1$ or $n=0$ used in the initial setup display the CE arrangement.
The intuitive reason is easy to explain. Simply to preserve  charge neutrality properly, the electronic
charge of the mobile electrons must rearrange itself at the interface to satisfy the Poisson equation. Then,
spontaneously, the electronic density at the interface is very close to $n=0.5$, by symmetry, 
a density conductive to the CE arrangement.
For more details please see Ref.~\cite{canted-CE}. Important is to note that while $Z=4$ is virtually a ``perfect'' CE state,
the other layers $Z=3$ and $5$ are not perfect. In particular, focus your attention on $Z=3$. In this
case, the typical zigzag chains of the CE state can be easily identified, but now the spins have acquired a clear {\it canting}
along the horizontal axis 
caused by the proximity to the FM layer $Z=2$. Namely, $Z=3$ is a compromise between the states in $Z=2$ and $Z=4$.

Thus, layer $Z=3$ is an exotic interpolation between the extreme cases of the FM layers of the
A-AFM state at one end and the CE state at the interface.  This canted
CE state is not present in the bulk phase diagram, neither theoretically nor experimentally. 
Thus, there is no evidence that the ``canted CE'' state exists in bulk form in real experiments, and, thus,
could only be created using artificial manganite superlattices. This example, consequently, belongs to the
same universality class as the previously discussed ``exotic'' skyrmions found when at the top and bottom of the
ensemble stripes in orthogonal directions were used. The results in Fig.~\ref{fig:newfrustration2} provide another example 
of the novel third frustration mechanism explained in previous sections, and it would be interesting for experimentalists
to confirm these intriguing theoretical predictions: new states of matter can exist in superlattices due a new third type of frustration.

%\begin{figure}
%\begin{center}
%\includegraphics[width=1\columnwidth, clip]{newfrustration3.jpg}
%\caption{Layer dependence of the spin structure
%factor $S({\bf k}$/$N$ (where $N$ is the number of sites of the cluster used) 
%for the optimized spin configurations shown in Fig. XX at
%several momenta of relevance. The black line is characteristic of A-AFM, the green line of a G-AFM,
%while the red and blue lines are characteristics of the CE state that {\it emerges} due to the transfer of charge
%at the interface.
%}
%\label{fig:newfrustration3}
%\end{center}
%\end{figure}

\subsection{Metallicity in superlattices induced by mixing insulating components}

In the same spirit as in the examples given above, we end this section by addressing a case where the components
of the superlattice are insulators when in bulk form, but when in  individually sufficiently {\it thin} superlattices
then  the ensemble becomes metallic, which is a remarkable result~\cite{LMO-SMO-elbio1}.

\begin{figure}
\begin{center}
\includegraphics[width=1\columnwidth, clip]{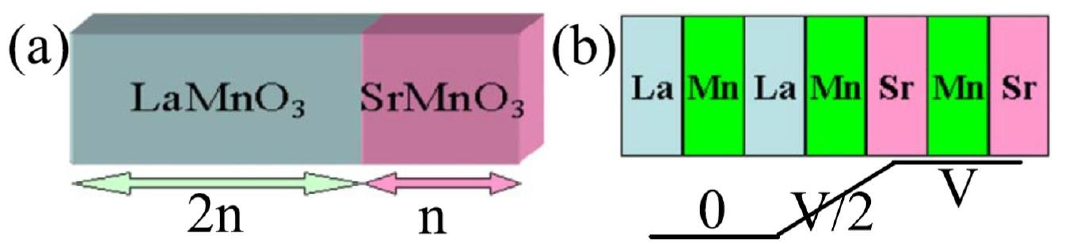}
\caption{Superlattice studied theoretically in Ref.~\cite{LMO-SMO-elbio1}. The components LaMnO$_3$ and SrMO$_3$ are in proportion 2:1.
Panel (a) are two thick examples of a superlattice. In this case global metallicity is not obtained. 
But for sufficiently thin cases, several of them one after the other, a global metallic sample is obtained because
of the inevitable transfer of electronic charge from SMO to LMO, or equivalently hole charge from LMO to SMO. Panel (b) represents
schematically an atomic plane-by-plane detail of the interface.  The sketch below panel (b) represents a potential $V$ that effectively
achieves the expected transfer of charge. Rather than using the far more challenging self-consistent Poisson equation procedure of the previous subsection, in this project
we used our physical intuition and introduced by hand the potential $V$, adjusting its specific value to reproduce the experimental
results. Fortunately, fine tuning of $V$ is not needed. 
Qualitatively, in practice adding $V$ is then almost the same as solving the far more involved Poisson equation. The benefit of using $V$ is that it allows  to study larger systems.
}
\label{fig:LMO-SMO-figure1}
\end{center}
\end{figure}

Intuitively, the story is simple to express. Suppose that experimentally 
a superlattice of LaMnO$_3$ and SrMnO$_3$ is created, as
illustrated in Fig.~\ref{fig:LMO-SMO-figure1}.
Both of these materials are known to be insulating when in bulk form. LaMnO$_3$ has both spin and orbital order, while SrMnO$_3$
simply has AFM spin staggered order. Then, we will create a superlattice involving both insulating materials in a proportion 2:1 (LMO:SMO).
The reason for this number is that the {\it average} electronic density then would be 2/3 of the electronic density of LMO alone,
and in the bulk phase diagram of manganites that places us in the {\it metallic} region, opening the possibility of metallicity emerging
from insulating components (see Refs.~\cite{elbio1-12,elbio1-13,elbio1-14,elbio1-15} and references therein). 
Namely, a new phase can be created from individual building blocks that do not have the properties of the 
resulting ensemble. Note that this result is not obvious. In the alloy form there is an intrinsic disorder due to the alloy
nature of the compound, while in superlattices at least some partial order, the one in each layer, is restored. Another important detail to remind the reader
is that LaMnO$_3$ while widely considered an A-AFM insulator, can also be a FM-insulator when affected by strain. In fact, LMO thin films
when grown on an appropriate susbtrated are FM (see Ref.~\cite{LMO-SMO-elbio1} and references therein). The orbital order in both cases remains staggered, as well as their insulating character.
In the following, we will describe some details of the calculations.

The Hamiltonian used in this study can be found in the original Ref.~\cite{LMO-SMO-elbio1}, but here it will be described qualitatively.
It contains the canonical double-exchange term characteristic of manganites, regulating the movement of $e_g$ electrons 
in the presence of the $t_{2g}$ spins, the latter widely assumed to be classical for simplicity in theoretical calculations. 
%Then, it also contains the Jahn-Teller distortions term involving oxygen displacements, 
%which are known to be very important in manganites that 
%display staggered lattice distortions. 
Moreover, the $t_{2g}$ spins interact among themselves with a strength that appears
to be weak in value, yet it produces important qualitative changes in the phase diagram. 
Finally, we have the combination of the global
chemical potential, used for the global electronic population, in addition to the  inhomogeneous potential $V$ 
as described in Fig.~\ref{fig:LMO-SMO-figure1} which qualitatively mimicks the effect of the long-range Coulomb interaction. This term
redistributes charge moving electrons from LMO to SMO (remember that La is in a valence $3+$ while Sr is in $2+$).
This effective potential replaces the use of the Poisson equation self-consistent procedure which is far more demanding
regarding computer time.

Overall, the resulting Hamiltonian is studied via Monte Carlo simulations on the classical localized $t_{2g}$
spins as well as on the lattice oxygen distortions (not quantized). In spite of the classical assumption, this task 
is still computationally challenging because any local change in the classical
variables requires a full rediagonalization of the fermionic sector. For technical details see Ref.~\cite{elbio1-14}.
Specifically, clusters of sizes $4 \times 4 \times 12$ were employed. The charge density, spin structure factor and conductivity
were calculated (see Ref.~\cite{LMO-SMO-elbio1} for details).

\begin{figure}
\begin{center}
\includegraphics[width=1\columnwidth, clip]{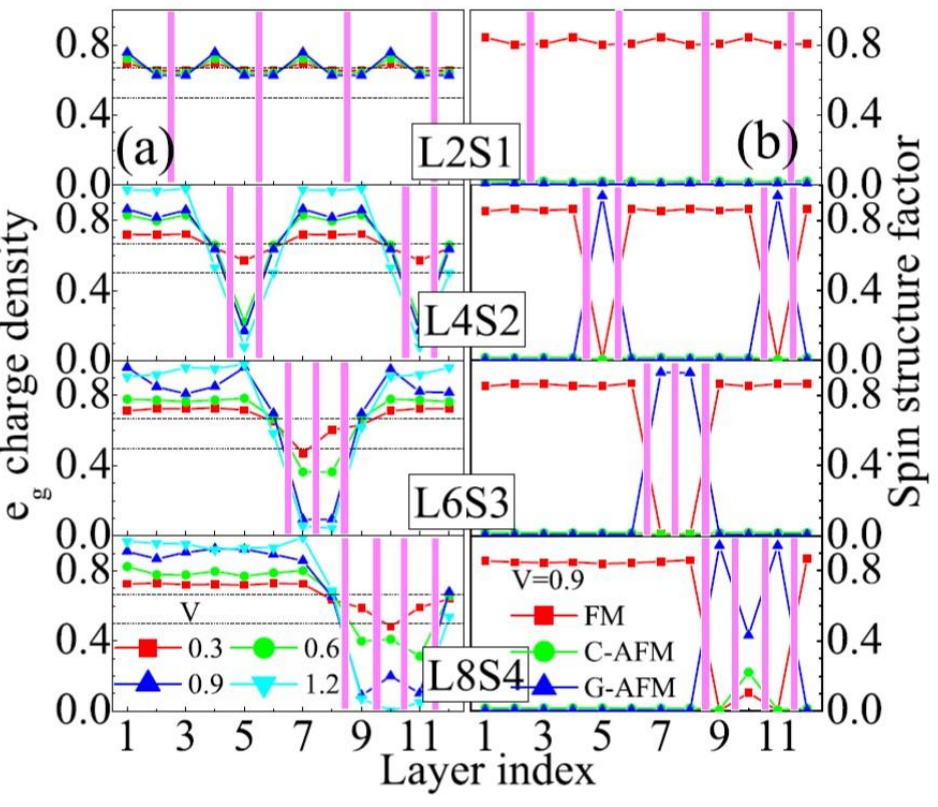}
\caption{(a) Charge modulation in the superlattices for different values of the parameter $V$ (see color convention 
at the bottom panel) that regulates the charge spreading, mimicking the effect of 
the long-range Coulomb interactions.The two horizontal lines denote 0.5 and
2/3. (b) In-plane spin structure factor for $V=0.9$. In both (a) and (b), pink
bars denote SrO layers in the (LMO)$_{2n}$/(SMO)$_n$ superlattice, while
LaO layers are not highlighted. Readers should focus on the two top panels of both (a) and (b). In the top panel of (a)
clear evidence of charge nearly {\it uniformly distributed at value 2/3} is given, together with a FM spin structure factor
as shown in the top panel of (b). This is the most clear case (very thin components) where the combination of two materials with
rather different properties produces an ensemble that is {\it nearly uniform}. For a more detailed description of the results
in the other panels, we refer the readers to the original Ref.~\cite{LMO-SMO-elbio1}.
}
\label{fig:LMO-SMO-figure2}
\end{center}
\end{figure}

In Fig.~\ref{fig:LMO-SMO-figure2}, details regarding the charge distribution, as well as the spin order, for different values
of the thickness $n$ are provided. For thin individual components of LMO and SMO, the charge can sufficiently spread to be almost
uniform at a value $n=2/3$, which in the bulk phase diagram corresponds to a FM metallic state. This is compatible with the values
found for the spin structure. As expected, in the same figure it is shown what occurs as the thickness of each component grows. Eventually, as this thickness becomes sufficiently
large, each component LMO or SMO maintains its original bulk properties, and no longer the ensemble behaves as a new material but
it is simply the obvious mixture of two insulators.

The issue of the global conductivity of the ensemble is more subtle and will not be reproduced here. Once again, for
details, please see Ref.~\cite{LMO-SMO-elbio1}. However, qualitatively it is clear that a metal to insulator transition can occur
in the theoretical simulation as the thickness of the components grows, as found experimentally~\cite{bhatta1,bhatta2,bhatta3,bhatta4}, providing yet one more example where from components with special characteristics, 
for sufficiently thin superlattices the global ensemble can behave entirely differently.

We also note that the emergent metallic behavior in superlattices of different insulating compounds is quite universal, not specific to the model considered here. 
In Ref.~\cite{Okamoto2004b}, the emergent metallic behavior is predicted for Mott-insulator/band-insulator heterostructures, relevant to the experimental results reported in Ref.~\cite{Ohtomo2002}. 
The many novel electronic phases discussed in the subsections of this section are indeed a manifestation of ``electronic reconstruction,'' i.e., emergence of electronic phases, that differ from the bulk electronic phases, in heterostructures introduced in Ref.~\cite{Okamoto2004}.

\section{Theoretical analysis of semi-Dirac bands in tetragonal transition-metal oxides}%S\lowercase{r}N\lowercase{b}O$_3$}
\label{sec:semidirac}

We now turn to the investigation of novel electronic states realized in strained TMOs. 
Specifically, we consider SrNbO$_3$ with the tetragonal symmetry, as experimentally studied in Ref.~\cite{Ok2021}. 
In contrast to the $e_g$ electron systems discussed in the previous sections, 
$t_{2g}$ electron systems are considered. 
As shown later, this allows one to include the SOC explicitly using the atomic-like form. 
By contrast, the atomic-like SOC in $e_g$ electron systems is activated when the local symmetry is lowered 
as discussed in Ref.~\cite{Xiao2011}.
%{\color{red}
After analyzing the electronic property of tetragonal SrNbO$_3$, 
we investigate the effect of magnetism using the effective model for tetragonal $t_{2g}$ TMOs. 
We will see this could lead to Weyl fermions and altermagnetism. 
%}

\subsection{Density Functional Theory analysis} 
To begin with, we analyze the electronic structure using {\it ab initio} density functional theory (DFT) to gain 
material specific information, which could be useful for constructing an effective model, as discussed later. 
For this purpose, we use the QUANTUM ESPRESSO code with the plane-wave pseudopotential method  \cite{QS}. 
We adopted the Perdew-Burke-Ernzerhof generalized gradient approximation exchange-correlation functional \cite{PBE}, 
a $16 \times 16 \times 16$ Monkhorst-Pack grid, and a plane wave energy cut-off of 600 eV.
Starting with the experimental lattice constants $a=b=5.518$~\r{A} and $c=8.280$~\r{A} \cite{Ok2021}, 
we performed a full optimization of the crystal structure using a force convergence criterion of $10^{-3}$~eV/\r{A}.

\begin{figure}
\begin{center}
\includegraphics[width=0.95\columnwidth, clip]{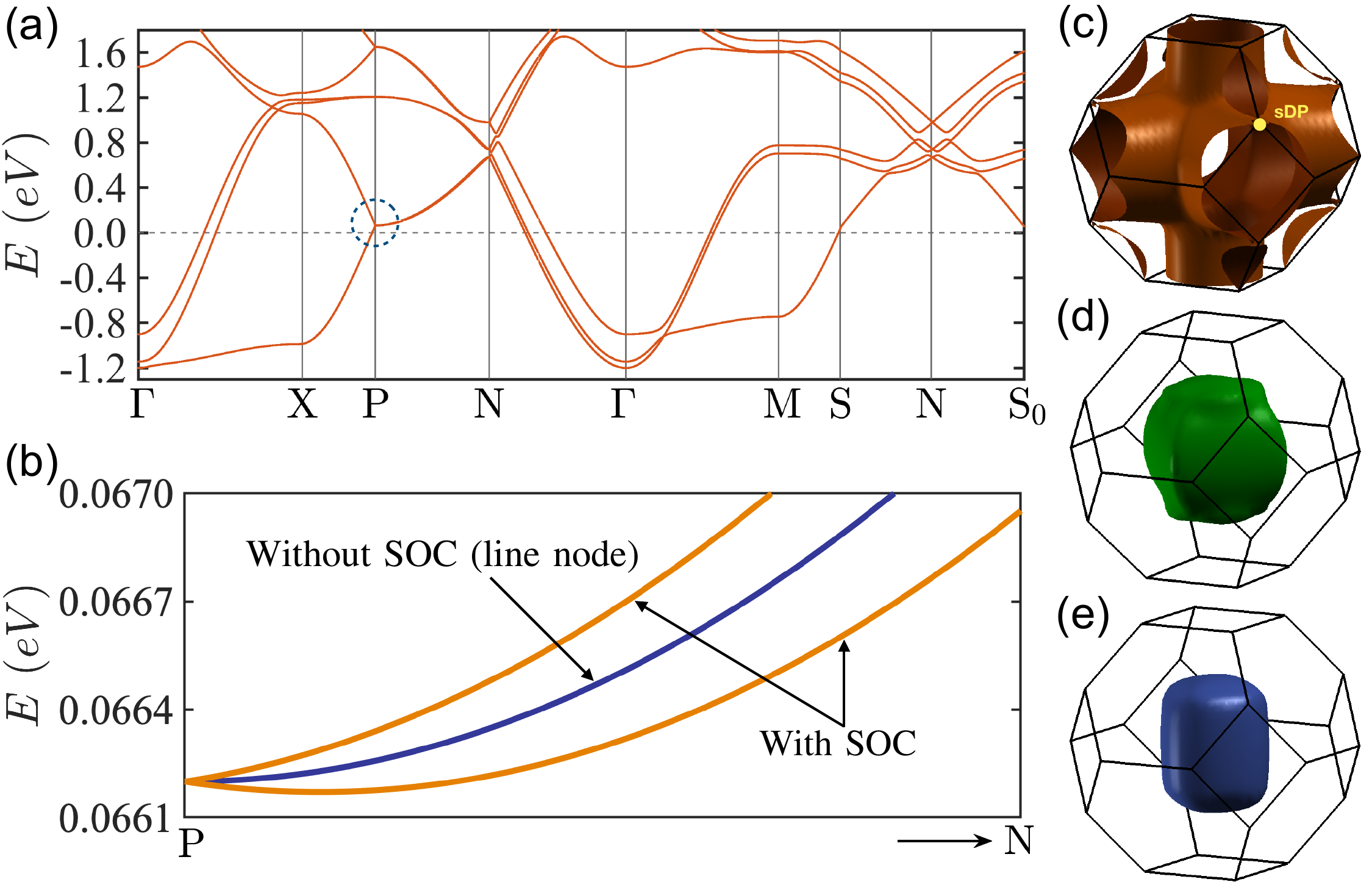}
\caption{(a) Band dispersion of tetragonal SrNbO$_3$ in the presence of spin-orbit coupling, 
through various high-symmetry momenta in the Brillouin zone. 
(b) Expanded view of the dispersion along the P-N direction, without and with spin-orbit coupling. 
(c)–(e) The three Fermi surfaces of the spin-orbit coupled $t_{2g}$ bands at the Fermi level. 
The semi-Dirac point (sDP) is denoted by the dashed circle at the P point in (a) and by the yellow dot in (c).
The figure is adopted from Ref.~\cite{Mohanta2021}.}
\label{fig:semidiracDFT}
\end{center}
\end{figure}

The band dispersion of tetragonal SrNbO$_3$ with finite SOC along high-symmetry momenta is shown in Fig.~\ref{fig:semidiracDFT}~(a). 
The $t_{2g}$​ orbitals of the Nb $4d$ orbital predominantly populate the Fermi level, similar to cubic SrNbO$_3$. 
Due to zone folding, the bands along the P-N momentum direction are fourfold degenerate without SOC, creating line nodes. 
When SOC is introduced, the degeneracy of these nodal lines is lifted everywhere except at high-symmetry points, as shown in Fig.~\ref{fig:semidiracDFT}~(b). 
The TRS and nonsymmorphic symmetry protect the semi-Dirac degeneracy at the P point at $E \approx 0.0662$~eV. 
Three other Dirac points appear at the N point above $E=0.5$~eV. 
Since these Dirac points are away from the Fermi level, they may not be accessible to transport. 
Therefore, we focus on the semi-Dirac point at the P point. 
However, replacing Nb with another transition-metal (TM) element, 
such as Mo, can tune the Fermi level close to the Dirac points at the N point. 
The Fermi surfaces of the spin-orbit coupled $t_{2g}$​ orbitals are depicted in Figs.~\ref{fig:semidiracDFT} (c)-(e). 
Due to the linear dispersions near the semi-Dirac point, a small chemical doping or gating can tune the Fermi level to the semi-Dirac point, 
creating the possibility of obtaining a quantized Berry phase and very high-mobility conduction.

%\subsection{tight-binding analysis of $t_{2g}$ electron systems with the tetragonal distortion}

\begin{figure}
\begin{center}
\includegraphics[width=1\columnwidth, clip]{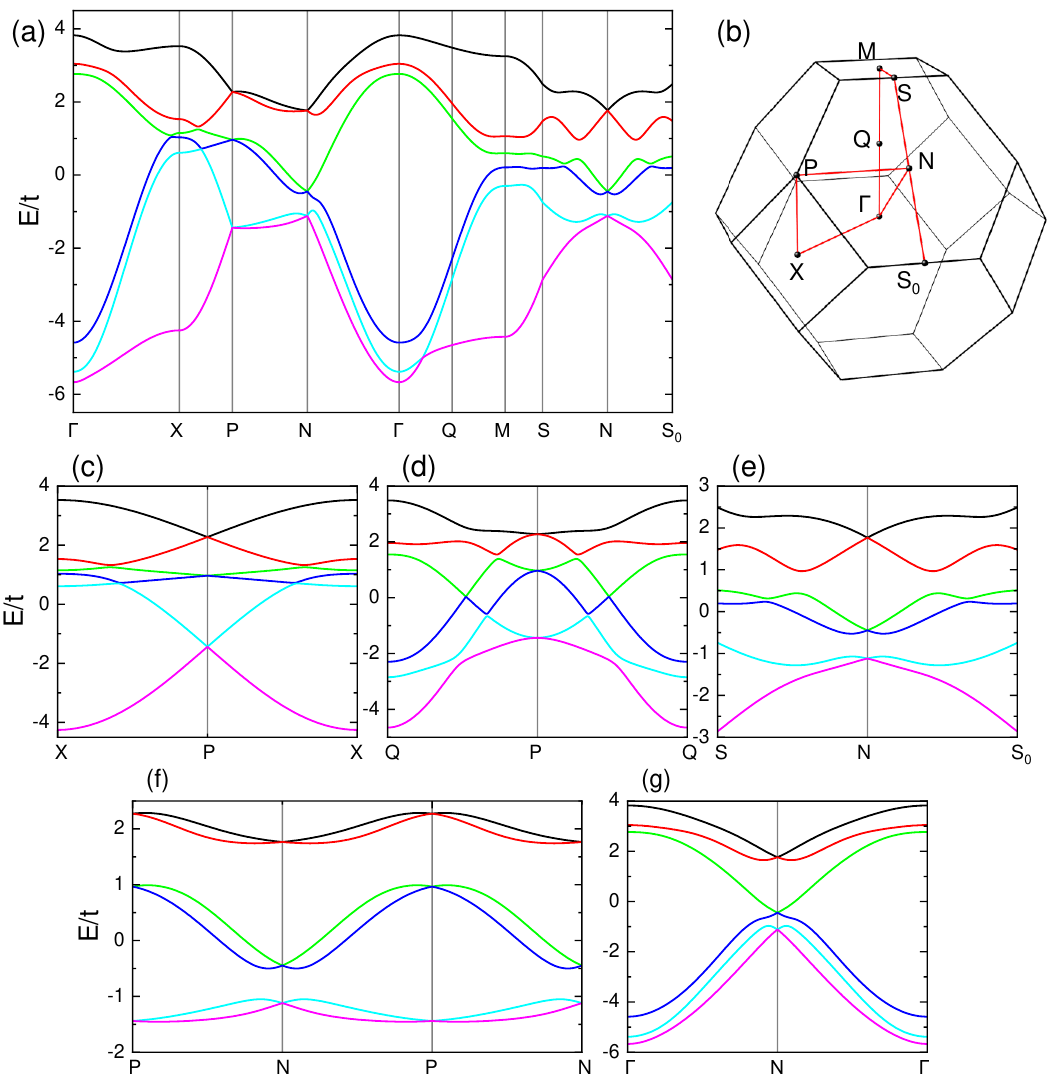}
\caption{(a) Band dispersion of the $t_{2g}$ tight-binding model 
through high-symmetry momenta in the Brillouin zone (b). 
(c)-(g) Expanded view of the dispersions across semi-Dirac and Dirac points at the P point and the N point. 
}
\label{fig:semidiracTB}
\end{center}
\end{figure}

\subsection{Tight-binding analysis}% of $t_{2g}$ electron systems with the tetragonal distortion}
To gain insight into the nature of semi-Dirac dispersions in $t_{2g}$ perovskite TMOs 
with the $I4/mcm$ crystalline symmetry and to carry out further analysis, 
we develop an effective tight-binding Hamiltonian as  
\begin{eqnarray}
H_{eff} = H_{TB} + H_{CFS} + H_{SOC}. 
\label{eq:Htetra}
\end{eqnarray}

For the sake of simplicity, we define the local orbitals as $\{a,b,c\}=\{d_{yz}, d_{zx}, d_{xy} \}$, 
and denote the hopping integral from orbital $\beta$ to $\alpha$ along the $\vect r$ direction as % = h \hat x + k \hat y + l \hat z$ direction as 
$t_{\vect r}^{\alpha \beta}$. %$t_{hkl}^{\alpha \beta}$. 
This leads to the following tight-binding Hamiltonian:
\begin{eqnarray}
H_{TB} =- \sum_{i, j} \sum_{\alpha, \beta, \sigma} 
 t_{hkl}^{\alpha \beta} c_{i \alpha \sigma}^\dag c_{j \beta \sigma}, 
\end{eqnarray}
where $c_{i \alpha \sigma}$ ($c_{i \alpha \sigma}^{\dag}$) is an annihilation (creation) operator of an electron on $\alpha$ orbital at position $\vect r_i$ with spin $\sigma$. 
$hkl$ denotes the hopping direction $\vect r_{ij}$ via $\vect r_{ij} = h \hat x + k \hat y + l \hat z$. 
%Typical hopping integrals are presented in Fig. \ref{fig:typicalhopping}. 
%The amplitude and the sign of $t_{hkl}^{\alpha \beta}$ are determined by the relative direction $h \hat x + k \hat y + l \hat z$ and the sign of two orbitals 
%and must  satisfy the underlying $I4/mcm$ symmetry. 
%
%In the undistorted perovskite structure, many hopping integrals disappear reflecting the symmetry 
%(positive and negative sign of $t_{2g}$ orbitals). 
%Among such hopping integrals, non-zero value of $t_{101}^{ac}$ and symmetry related hopping 
%as well as finite spin-orbit coupling
%are found to lift the band degeneracy of nodal lines along the P-N direction,  
%creating semi-Dirac dispersions at the P point and Dirac dispersions at the N point. 
%
Under the tetragonal distortion, the crystal field splitting is allowed between $a$, $b$, and $c$ as 
\begin{equation}
H_{CFS} = \sum_{i, \sigma} \bigl\{ \varepsilon_{ab} (c_{i a \sigma}^\dag c_{i a \sigma} +c_{i b \sigma}^\dag c_{i b \sigma}) 
+ \varepsilon_c c_{i c \sigma}^\dag c_{i c \sigma} \bigr\}, 
\end{equation}
where $a$ and $b$ remain degenerate in the $I4/mcm$ symmetry. 
Finally, the spin-orbit coupling is included using the atomic-like form given by  
\begin{equation}
H_{SOC} %= \bigl(\lambda \vec l \cdot \vec s \bigr)_{t_{2g}} 
= \frac{\lambda}{2} \sum_i \sum_{\alpha, \beta, \gamma} 
\sum_{\sigma, \sigma'}
{\rm i} \varepsilon_{\alpha \beta \gamma}  c_{i \alpha \sigma}^\dag \sigma_{\sigma \sigma'}^\gamma c_{i \beta \sigma'}, 
\end{equation}
where %$(\ldots)_{t_{2g}}$ indicates the projection onto the $t_{2g}$ subspace, 
%$\alpha$, $\beta$, and $\gamma$ are the orbital indices, $a$, $b$, and $c$, 
$\sigma^\gamma$ is the Pauli matrix, and $\gamma = a, b$, and $c$ should be interpreted as $x,y$, and $z$, respectively. 
%$\varepsilon_{\alpha \beta \gamma}$ is the Levi-Civita antisymmetric tensor. 

A typical dispersion relation of this theoretical model is plotted in Fig.~\ref{fig:semidiracTB}. 
For the hopping parameters, we considered 
$t_{010,001}^{aa}=t_{100,001}^{bb}=t_{100,010}^{cc}=t$,
$t_{100}^{aa}=t_{010}^{bb}=t_{001}^{cc}=0.1 t$, % t3
$t_{001}^{ab'} = - t_{001}^{ba'} = 0.2 t$, % tb
$t_{110}^{cc} = t_{011}^{aa} = t_{101}^{bb} = 0.2 t$, % t2
$t_{1\bar1 0}^{aa} = t_{11 0}^{bb} =0.2 t$, % ta
$t_{110}^{ab} = - t_{1\bar10}^{ab} = 0.2 t$, % t4
$t_{101}^{ac} = t_{10\bar1}^{ac}= 0.2 t$. % t4 
Here, some of the off-diagonal hopping parameters along the $z$ direction explicitly depend on the sublattice index as indicated by $t_{001}^{ab'}$ and $t_{001}^{ba'}$. 
Other hopping parameters are related to the above parameters by the ${\cal C}_{4z}$ rotational symmetry or zero. 
Other parameters are $\varepsilon_c=0.2t$ and $\lambda =0.5 t$. 
Note that this tight-binding dispersion relation qualitatively reproduces the DFT dispersions, 
in particular the semi-Dirac dispersions at the P point and the linear Dirac dispersions at the N point. 

Detailed analyses on the fourfold degeneracy at the N and P points, i.e. Dirac band crossing, are provided in Ref.~\cite{Mohanta2021}. 
Glide ${\cal G}_{xy,x \bar y}^z$ symmetries protect the spin degeneracy at the N point, 
where $G_{xy,x \bar y}^z$ consist of the mirror reflection about $k_x \mp k_y=0$ planes and half lattice translation along the $z$ direction.

\begin{figure}
\begin{center}
\includegraphics[width=1\columnwidth, clip]{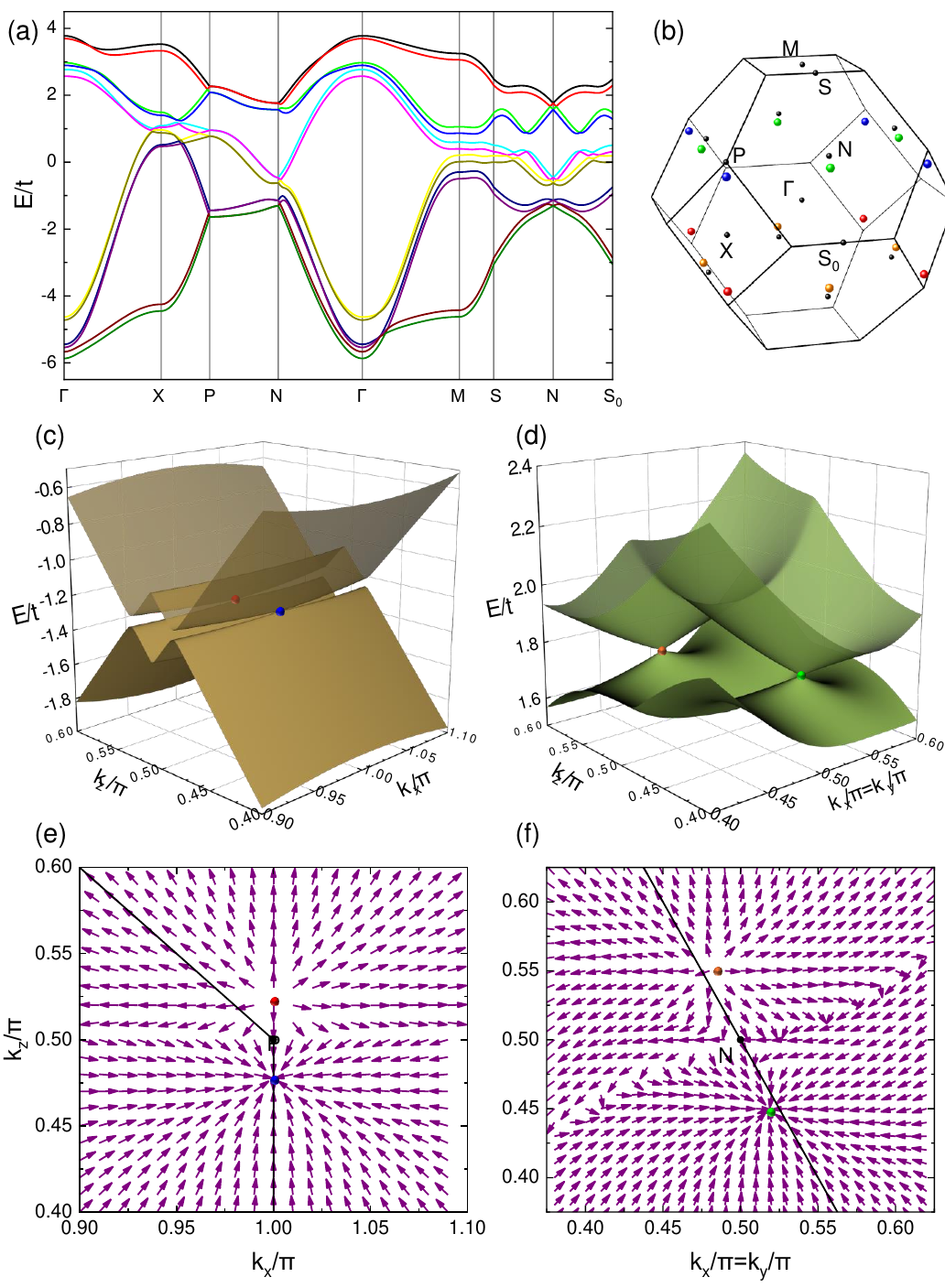}
\caption{(a) Band dispersion of the $t_{2g}$ tight-binding model with the uniform Zeeman field along the $z$ direction. 
Eight sets of Weyl points appear near the P or the N points as indicated in (b). 
(c) and (d) show dispersions of the second- and third-lowest bands near the P point with $k_y=0$ and 
the second- and third-highest bands near the N point with $k_x=k_y$, respectively. 
Red and Blue (orange and green) dots form pairs of Weyl points with the opposite chirality, 
as indicated by the source or the sink of the normalized Berry curvature vectors 
${\mathbf \Omega}_m(\vect k)/|{\mathbf \Omega}_m(\vect k)|$
shown in (e) near the P point and (f) near the N point. 
}
\label{fig:weyl}
\end{center}
\end{figure}

Breaking the TRS in a Dirac semimetal splits the spin degeneracy, leading to a magnetic semiconductor or a Weyl semimetal \cite{Armitage2018}. 
Such a possibility in tetragonal SrNbO$_3$ was discussed in Ref.~\cite{Mohanta2021}. 
Here, we demonstrate this behavior using our effective model. 
We introduce a uniform Zeeman field along the $z$ direction, represented by 
$-\sum_{i, \alpha} h_{uni} \sigma c_{i \alpha \sigma}^\dag c_{i \alpha \sigma}$ with $h_{uni} = 0.2 t$.
The resulting band dispersion is shown in  Fig.~\ref{fig:weyl}~(a), where Dirac band crossings originally seen at the P point and the N point split, 
leading to Weyl band crossings. 
Due to spatial inversion ($\cal P$) symmetry, the Weyl points appear as pairs at the same energy but with opposite chirality between
$\vect k$ and $-\vect k$, as depicted in Fig.~\ref{fig:weyl}~(b). 
Near the P point, a pair of Weyl points between the second- and third-lowest bands appear at 
$(\pi,0,\pi/2 \pm \delta_P) $ with $\delta_P \sim 0.02 \pi$ as shown in Fig.~\ref{fig:weyl}~(c). 
Near the N point, a pair of Weyl points between the second- and third-highest bands appear at $(\pi/2 \pm \delta_{N,1}, \pi/2 \pm \delta_{N,1} ,\pi/2 \mp \delta_{N,2}) $ with 
$\delta_{N,1} \sim 0.02 \pi$ and $\delta_{N,2} \sim 0.05 \pi$, as shown in Fig.~\ref{fig:weyl}~(d).

To check the chirality of the Weyl points, 
we analyzed the Berry curvature vector given by 
${\mathbf \Omega}_{m \vect k} = (\Omega^x_{m \vect k }, \Omega^y_{m \vect k }, \Omega^z_{m \vect k })$. 
$\Omega_{m \vect k}^\alpha$ are computed using Eq.~(\ref{eq:BCV}). 
As shown in Figs.~\ref{fig:weyl}~(e) and (f), ${\mathbf \Omega}_{m \vect k}$ have 
sources (indicated by red or green dots) and sinks (blue or green dots), corresponding to the positive and negative chirality, respectively. 

In TMO heterostructures, the time-reversal ($\cal T$) symmetry breaking (and additional symmetry breaking) could be induced  spontaneously by correlations or 
externally by using a magnetic substrate.

\subsection{Possible altermagnetism}% in $t_{2g}$ systems with the tetragonal distortion} 
%{\it Possible altermagnetism in $t{2g}$ systems with the tetragonal distortion:} 
%
In our discussions so far, we have established that lattice distortion in TMO oxides can be controlled by strain. 
Specifically, TM-O-TM bonds are distorted under compressive strain, resulting in the loss of the inversion center between NN sites. 
Tetragonal TMOs offer an opportunity to explore novel electronic states in such an environment. 
When additional N{\'e}el-type AFM ordering is present, momentum-dependent spin splitting can occur in the electron band structure 
\cite{Noda2016,Okugawa2018,Naka2019,Hayami2019,Hayami2020,Naka2021}. 
This new class of magnetism, %{\color{red} 
as explained in the introduction, is called %} 
altermagnetism \cite{Smejkal2022a,Smejkal2022b}, 
%{\color{red} 
and it could %} 
give rise to various phenomena potentially useful for spintronics applications 
with itinerant electrons, such as spin filtering. 
Here, we demonstrate that altermagnetism indeed appears in our model for a tetragonal $t_{2g}$​ system.

\begin{figure}
\begin{center}
\includegraphics[width=1\columnwidth, clip]{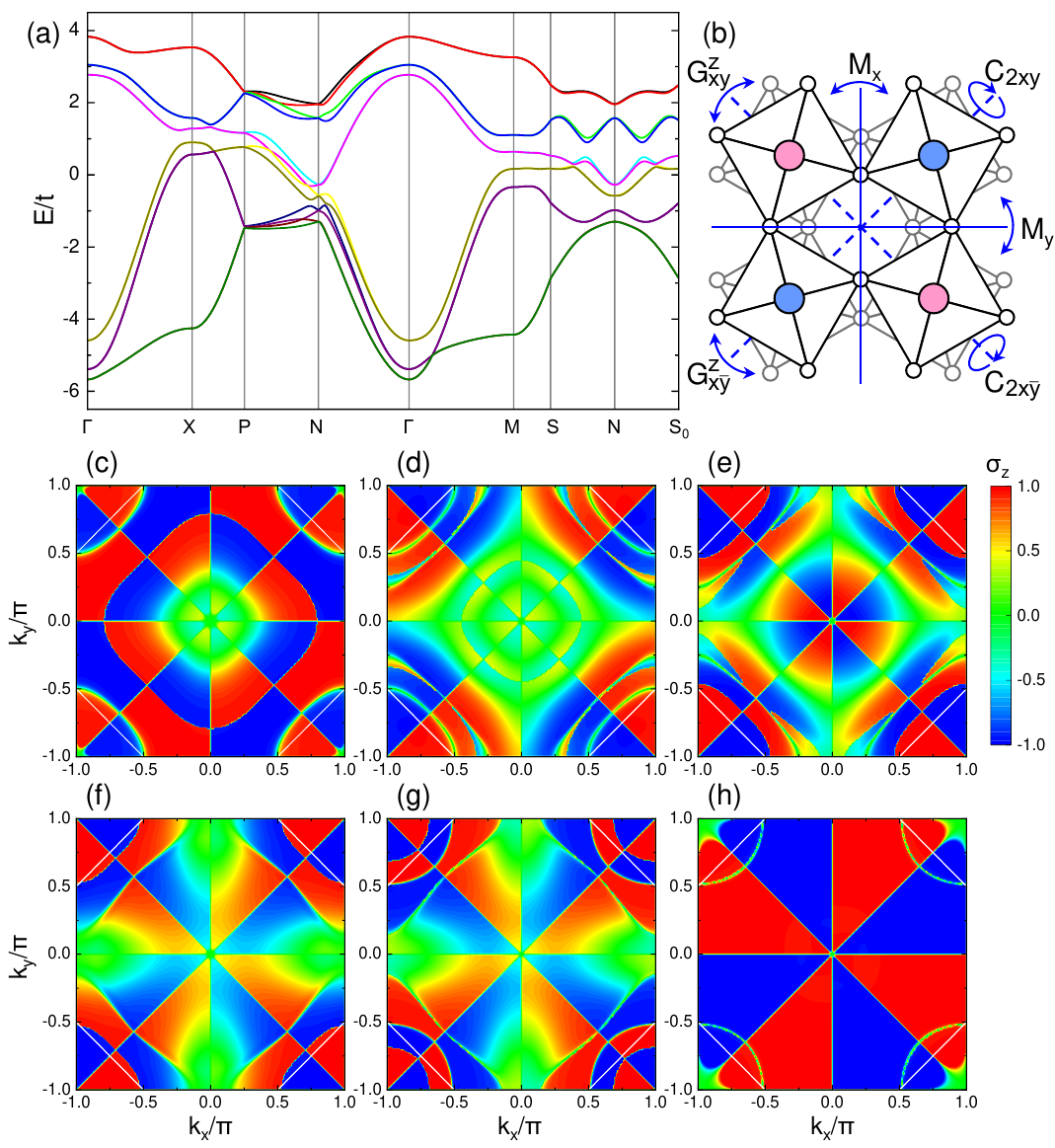}
\caption{(a) Band dispersion of the $t_{2g}$ tight-binding model with the alternating TRS breaking field along the $z$ direction. 
(b) Magnetic symmetry. Red and blue sites have opposite spin polarizations along the $z$ direction. 
Mirror (${\cal M}_{x,y}$) symmetries protect the spin degeneracy along $k_{y,x}=0$, 
two-fold rotation (${\cal C}_{2 xy, 2x \bar y}$) symmetries protect the spin degeneracy at $k_x=\pm k_y$ with $k_z=0,\pi$, and 
glide ${\cal G}_{xy,x \bar y}^z$ symmetries protect the spin degeneracy at the N point.   
Momentum dependent spin splitting of the $(2n-1)$-th band from the top is shown in 
(c) $n=1$, (d) $n=2$, (e) $n=3$, (f) $n=4$, (g) $n=5$, and (h) $n=6$ at $k_z=0$. 
Brillouin zone boundaries are indicated by white lines.
The spin splitting of the $2n$-th band (not shown) is opposite to that of the $(2n-1)$-th band. 
}
\label{fig:altermagnetism}
\end{center}
\end{figure}

To this end, we introduce a staggered effective magnetic field $-\sum_{i, \alpha} h_{stagg.} \sigma c_{i \alpha \sigma}^\dag c_{i \alpha \sigma} {\rm e}^{{\rm i} \vect Q \cdot \vect r_i}$ with 
$h_{stagg.} = 0.2 t$ and $\vect Q = (\pi,\pi,\pi)$ to Eq.~(\ref{eq:Htetra}).  
The resulting band dispersion is presented in Fig.~\ref{fig:altermagnetism}~(a). 
This system has mirror ($\cal M$), rotation ($\cal C$), and glide symmetries ($\cal G$) as shown in (b). 
These symmetries protect the spin degeneracy along high-symmetry $\vect k$ points. 
Most notably, the momentum-dependent spin splitting has the $g$-wave symmetry as shown in (c)-(h), 
reflecting the ${\cal C}_{4z}$ rotational symmetry. 

While some TMOs have been predicted to host altermagnetism \cite{Smejkal2022a,Smejkal2022b}, 
many candidate oxide compounds are insulating with a large gap. 
Strain engineering demonstrated in Ref.~\cite{Ok2021} could become extremely useful to induce altermagnetism in
conventional AFM metals.

\section{Second-order topological superconducting phase in an oxide trilayer}
\label{sec:tsc}

In this section, we discuss 
heteroengineering of lattice geometry in 
heterostructures of cubic perovskite TMOs grown along an ``unconventional'' direction. 
In Ref.~\cite{Xiao2011}, we proposed bilayers of TMOs grown along the [111] crystallographic direction as a possible route to realize topological insulating states. 
Such a bilayer forms a buckled honeycomb lattice, 
and the cubic crystalline symmetry is reduced to the trigonal symmetry. 
The possibility of topological superconducting states in such [111] bilayers was discussed in Ref.~\cite{Okamoto2013}. 

\begin{figure*}
\begin{center}
\includegraphics[width=2\columnwidth, clip]{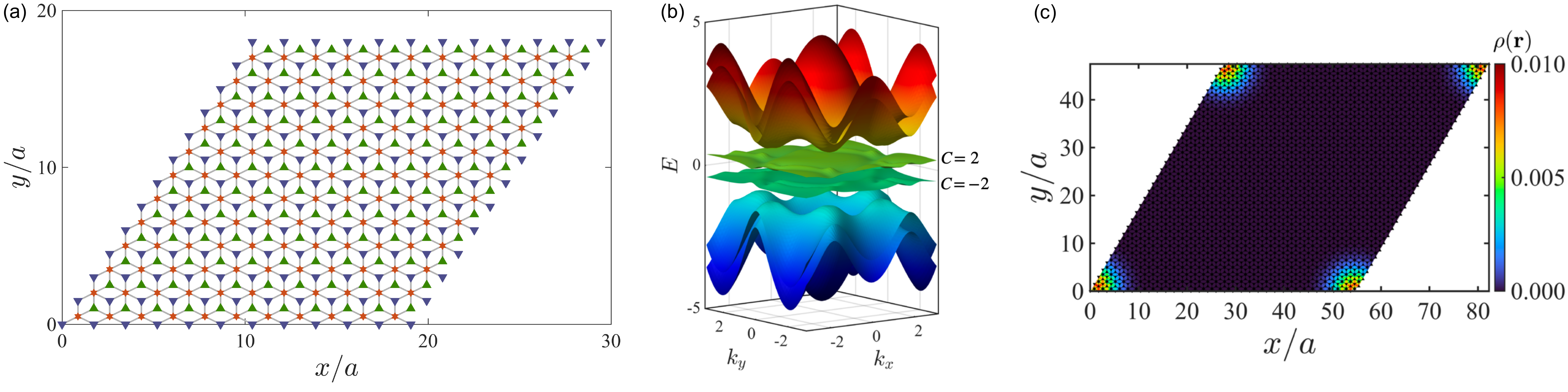}
\caption{(a) Schematic picture of the dice lattice, showing three inequivalent sites in each unit cell. 
(b) Electronic bands of the dice lattice in the presence of a spin-orbit coupling and a magnetic field, showing nearly-flat topological bands with Chern number $C=\pm 2$, close to the Fermi level. 
(c) Local density of states $\rho(\vect r)$ (in arbitrary units) in the topological superconducting phase obtained on a $32\times32$ dice lattice, showing the corner-localized Majorana bound states at zero energy. 
For a detailed description, we refer the readers to Ref.~\cite{Dice_CommPhys2023}.
}
\label{fig:Dice_Majorana}
\end{center}
\end{figure*}

Now we consider trilayers of cubic TMOs, such as SrTiO$_3$/[SrIrO$_3$]$_3$/SrTiO$_3$. 
As shown in Fig.~\ref{fig:Dice_Majorana}~(a), such a trilayer forms the dice lattice. 
Due to quantum interference, such a lattice exhibits a pair of flat electronic bands at the Fermi level. 
The magnetism originating from the strongly-correlated $d$ electrons of the Ir atoms and the Rashba spin-orbit interaction which may arise in this trilayer oxide geometry due to broken inversion symmetry induce an energy splitting of the flat bands, as shown in Fig.~\ref{fig:Dice_Majorana}(b). 
These split nearly-flat bands exhibit Chern numbers $C=\pm 2$~\cite{Wang_PRB2011,Soni_PRB2020}. 
It is known that when doped with elements such as Nb, SrTiO$_3$ undergoes a superconducting transition~\cite{Pfeiffer_PhysLett1969,Binnig_PRL1980}. 
It was shown by us that a second-order topological superconducting phase can be realized in this trilayer oxide geometry when the lower topological flat band is populated at the Fermi level~\cite{Dice_CommPhys2023}. 

%--------------------------------------------------------------------------
%--------------------------------------------------------------------------
A hallmark of the second-order topological superconducting phase is the localization of zero-energy Majorana bound states at the lattice corners. 
The corner-localized zero-energy Majorana bound states in this oxide dice lattice is shown in Fig.~\ref{fig:Dice_Majorana}~(c), where the local density of states $\rho(\vect r)$, corresponding to the zero-energy quasiparticle state in the topological superconducting phase, is plotted on the two-dimensional lattice. 
This finding establishes a close connection between the topology of electronic bands in the normal state and the localization of the zero-energy Majorana bound states. 
This theoretical finding also calls for experimental verification of the second-order topological superconductivity, which is typically rare in real materials, in a simple easy-to-fabricate oxide trilayer platform.

\section{Concluding remarks} 

\begin{figure*}
\begin{center}
\includegraphics[width=1.7\columnwidth, clip]{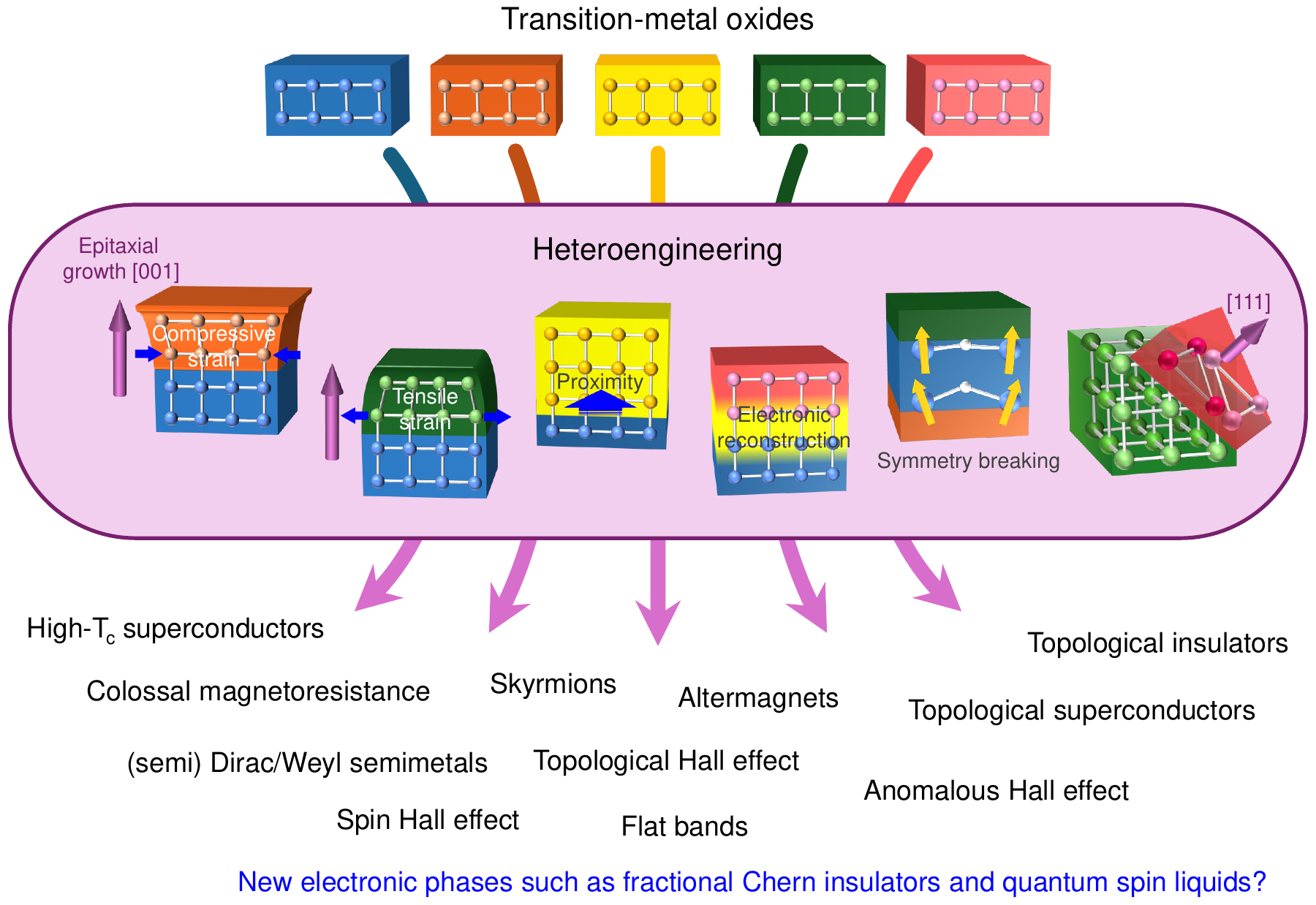}
\caption{Illustration of heteroengineering of transition-meal oxides. 
By combining different TMOs, one can control strain fields, superconducting or magnetic proximity effects, 
the leakage of electron charge leading to electronic reconstruction, 
and the lattice symmetry by breaking crystal symmetry or growing along an unconventional direction such as [111] crystallographic direction. 
This could lead to novel electronic phases or phenomena not observed in bulk crystalline systems. 
}
\label{fig:heteroengineering}
\end{center}
\end{figure*}

In this article, we presented selected portions of our theoretical work on transition-metal oxides with strong electron-electron interactions and relativistic spin-orbit coupling, when in the form of heterostructures. 
As demonstrated in this article, as well as in the experimental and theoretical references, ``heteroengineering'' of transition-metal oxides 
is extremely useful to control the lattice symmetry and geometry as illustrated in Fig.~\ref{fig:heteroengineering}.  
This could lead to novel electronic phases {\it not} observed in bulk crystalline systems. 
Our predictions include a novel form of skyrmion lattices, spin and orbital ordering, Weyl fermions, altermagnetism, and higher-order topological superconductivity, among many others. 
We hope that this article will stimulate further research, both theoretical and experimental, 
contributing to more efficient electronics, spintronics, and quantum computing.

\section*{Acknowledgment}
This work was supported by the U.S. Department of Energy, Office of Science, Basic Energy Sciences, Materials Sciences and Engineering Division. 
We thank S. Dong, E. Skoropata, and J. M. Ok for fruitful discussions and collaborations.


\begin{thebibliography}{*}

\bibitem{Imada1998}M. Imada, A. Fujimori, and Y. Tokura, 
Metal-insulator transitions. 
Rev. Mod. Phys. {\bf 70}, 1039 (1998). 

\bibitem{Tokura2000}Y. Tokura and N Nagaosa, 
Orbital physics in transition-metal oxides, 
Science {\bf 288}, 462 (2000). 

\bibitem{Ohtomo2002}A. Ohtomo, D. A. Muller, J. L. Grazul, and H. Y. Hwang, 
Artificial charge–modulation in atomic-scale perovslite titanate superlattices, 
Nature (London) {\bf 419}, 378 (2002). 

\bibitem{Okamoto2004}S. Okamoto and A. J. Millis, 
Electronic reconstruction at an interface between a Mott insulator and a band insulator,
Nature (London) {\bf 428}, 630; 
Theory of Mott insulator-band insulator heterostructure,
Phys. Rev. B {\bf 70}, 075101 (2004).

\bibitem{Lee2005}H. N. Lee, H. M. Christen, M. F. Chisholm, C. M. Rouleau, and D. H. Lowndes, 
Strong polarization enhancement in asymmetric three-component ferroelectric superlattices
Nature (London) {\bf 433}, 395 (2005). 

\bibitem{Sefrioui2003}Z. Sefrioui, D. Arias, V. Pe{\~n}a, J. E. Villegas, M. Varela, P. Prieto, C. Le{\'o}n, J. L. Martinez, and J. Santamar{\'i}a, 
Ferromagnetic/superconducting proximity effect in La$_{0.7}$Ca$_{0.3}$MnO$_3$/YBa$_2$Cu$_3$O$_{7-\delta}$ superlattices, 
Phys. Rev. B {\bf 67}, 214511 (2003). 

\bibitem{Nemes2008}N. M. Nemes, M. Garc{\'i}a-Hern{\'a}ndez, S. G. E. te Velthuis, A. Hoffmann, C. Visani, J. Garcia-Barriocanal, V. Pe{\~n}a, D. Arias, Z. Sefrioui, C. Leon, and J. Santamar{\'i}a, 
Origin of the inverse spin-switch behavior in manganite/cuprate/manganite trilayers, 
Phys. Rev. B {\bf 78}, 094515 (2008). 

\bibitem{Chakhalian2006}J. Chakhalian, J. W. Freeland, G. Srajer, J. Strempfer, G. Khaliullin, J. C. Cezar, T. Charlton, R. Dalgliesh, C. Bernhard, G. Cristiani, H.-U. Habermeier, and B. Keimer, 
Magnetism at the interface between ferromagnetic and superconducting oxides, 
Nat. Phys. {\bf 2}, 244 (2006). 

\bibitem{Salafranca2010}J. Salafranca and S. Okamoto, 
Unconventional Proximity Effect and Inverse Spin-Switch Behavior in a Model Manganite-Cuprate-Manganite Trilayer System, 
Phys. Rev. Lett. {\bf 105}, 256804 (2010). 

\bibitem{Okamoto2010}S. Okamoto, 
Magnetic interaction at an interface between manganite and other transition-metal oxides,
Phys. Rev. B {\bf 82}, 024427 (2010).

\bibitem{Yu2010a} P. Yu, J.-S. Lee, S. Okamoto, M. D. Rossell, M. Huijben, C.-H. Yang, Q. He, J. X. Zhang, S.Y. Yang, M. J. Lee, Q. M. Ramasse, R. Erni, Y.-H. Chu, D. A. Arena, C.-C. Kao, L.W. Martin, and R. Ramesh,
Interface ferromagnetism and orbital reconstruction in BiFeO$_3$-La$_{0.7}$Sr$_{0.3}$MnO$_3$ heterostructure,
Phys. Rev. Lett. {\bf 105}, 027201 (2010).


\bibitem{Nichols2016}J. Nichols, X. Gao, S. Lee, T. L. Meyer, J. W. Freeland, V. Lauter, D. Yi, J. Liu, D. Haskel, J. R. Petrie, E.-J. Guo, A. Herklotz, D. Lee, T. Z. Ward, G. Eres, M. R. Fitzsimmons, and H. N. Lee, 
Emerging magnetism and anomalous Hall effect in iridate-manganite heterostructures, 
Nat. Commun. {\bf 7}, 12721 (2016).

\bibitem{Okamoto2017}S. Okamoto, J. Nichols, C. Sohn, S. Y. Kim, T. W. Noh, and H. N. Lee, 
Charge transfer in iridate-manganite superlattices, 
Nano Lett. {\bf 17}, 2126 (2017).



\bibitem{Dzyaloshinskii1957}I. E. Dzyaloshinskii, Thermodynamic theory of ``Weak'' ferromagnetism in antiferromagnetic substances, 
Sov. Phys. JETP. {\bf 5}, 1259 (1957).

\bibitem{Moriya1960}T. Moriya, 
New mechanism of anisotropic superexchange interaction,  
Phys. Rev. Lett. {\bf 4}, 228 (1960).



\bibitem{Bogdanov2001}A. N. Bogdanov and U. K. R{\"o}{\ss}ler, 
Chiral Symmetry Breaking in Magnetic Thin Films and Multilayers, 
Phys. Rev. Lett. {\bf 87}, 037203 (2001). 

\bibitem{Muhlbauer2009}S. M{\"u}hlbauer, B. Binz, F. Jonietz, C. Pfleiderer, A. Rosch, A. Neubauer, R. Georgii, P. B{\"o}ni, 
Skyrmion Lattice in a Chiral Magnet, 
Science {\bf 323}, 915 (2009). 

\bibitem{Yu2010b}X. Z. Yu, Y. Onose, N. Kanazawa, J. H. Park, J. H. Han, Y. Matsui, N. Nagaosa, and Y. Tokura, 
Real-space observation of a two-dimensional skyrmion crystal, 
Nature (London) {\bf 465}, 901 (2010). 

\bibitem{Tokura2021}Y. Tokura and N. Kanazawa, 
Magnetic Skyrmion Materials, 
Chem. Rev. {\bf 121}, 2857 (2021). 

% THE 

\bibitem{Taguchi2001}Y. Taguchi, Y. Oohara, H. Yoshizawa, N. Nagaosa, and Y. Tokura, 
Spin Chirality, Berry Phase, and Anomalous Hall Effect in a Frustrated Ferromagnet, 
Science {\bf 291}, 2537 (2001). 


\bibitem{Matsuno2016}J. Matsuno, N. Ogawa, K. Yasuda, F. Kagawa, W. Koshibae, N. Nagaosa, Y. Tokura, M. Kawasaki, 
Interface-driven topological Hall effect in SrRuO$_3$-SrIrO$_3$ bilayer, 
Sci. Adv. {\bf 2}, e1600304 (2016).


\bibitem{Kan2018}D. Kan, T. Moriyama, K. Kobayashi, and Y. Shimakawa,
Alternative to the topological interpretation of the transverse resistivity anomalies in SrRuO$_3$,
Phys. Rev. B {\bf 98}, 180408(R) (2018). 

\bibitem{Kimbell2020}G. Kimbell, P. M. Sass, B. Woltjes, E. K. Ko, T. W. Noh, W. Wu, and J. W. A. Robinson,
Two-channel anomalous Hall effect in SrRuO$_3$,
Phys. Rev. Materials {\bf 4}, 054414 (2020). 

\bibitem{Roy2023}P. Roy, A. Carr, T. Zhou, B. Paudel, X. Wang, D. Chen, K. T. Kang, A. Pateras, Z. Corey, S. Lin, J.-X. Zhu, M. V. Holt, J. Yoo, V. Zapf, H. Zeng, F. Ronning, Q. Jia, A. Chen, 
Origin of Topological Hall-Like Feature in Epitaxial SrRuO3 Thin Films,
Adv. Electron. Mater. {\bf 9}, 2300020 (2023). 


\bibitem{Li2019}Y. Li, L. Zhang, Q. Zhang, C. Li, T. Yang, Y. Deng, L. Gu, and D. Wu, 
Emergent topological Hall effect in La$_{0.7}$Sr$_{0.3}$MnO$_3$/SrIrO$_3$ Heterostructures,
 ACS Appl. Mater. Interfaces {\bf 11}, 21268 (2019). 

\bibitem{Skoropata2020}E. Skoropata, J. Nichols, J. M. Ok, R. V. Chopdekar, E. S. Choi, A. Rastogi, C. Sohn, X. Gao, S. Yoon, T. Farmer, R. Desautels, Y. Choi, D. Haskel, J. W. Freeland, S. Okamoto, M. Brahlek, and H. N. Lee, 
Interfacial tuning of chiral magnetic interactions for large topological Hall effects in LaMnO$_3$/SrIrO$_3$ heterostructures,
Sci. Adv. {\bf 6}, eaaz3902 (2020).

\bibitem{vesta}K. Momma and F. Izumi, ``VESTA 3 for three-dimensional visualization of crystal, volumetric and morphology data,'' J. Appl. Crystallogr. {\bf 44}, 1272  (2011).

% AHE in SrIrO3

\bibitem{Yoo2021}M.-W. Yoo, J. Tornos, A. Sander, L.-F. Lin, N. Mohanta, A. Peralta, D. Sanchez-Manzano, F. Gallego, D. Haskel, J. W. Freeland, D. J. Keavney, Y. Choi, J. Strempfer, X. Wang, M. Cabero, H. B. Vasili, M. Valvidares, 
J. M. Gonzalez-Calbet, A. Rivera, C. Leon, S. Rosenkranz, M. Bibes, A. Barthelemy, A. Anane, E. Dagotto, S. Okamoto, S. G. E. te Velthuis, J. Santamaria, and J. E. Villegas,
Large intrinsic anomalous Hall effect in SrIrO$_3$ induced by magnetic proximity effect,
Nat. Commun. {\bf 12}, 3283 (2021).


% Topological insulator, semimetal, etc

\bibitem{Murakami2003}S. Murakami, N. Nagaosa, and S.-C. Zhang, 
Dissipationless Quantum Spin Current at Room Temperature, 
Science {\bf 301},1348 (2003). 

\bibitem{Kane2005}C. L. Kane and E. J. Mele, 
$Z_2$ Topological Order and the Quantum Spin Hall Effect, 
Phys. Rev. Lett. {\bf 95}, 146802 (2005). 

\bibitem{Bernevig2006a}B. A. Bernevig and S.-C. Zhang, 
Quantum Spin Hall Effect, 
Phys. Rev. Lett. {\bf 96}, 106802 (2006). 

\bibitem{Bernevig2006b}B. A. Bernevig, T. L. Hughes, S.-C. Zhang, 
Quantum spin Hall effect and topological phase transition in HgTe quantum wells
Science {\bf 314}, 1757 (2006). 


\bibitem{Fu2007}L. Fu, C. L. Kane, and E. J. Mele, 
Topological Insulators in Three Dimensions, 
Phys. Rev. Lett. {\bf 98}, 106803 (2007). 

\bibitem{Konig2007}M. K{\"o}nig, S. Wiedmann, C. Br{\"u}ne, A. Roth, H. Buhmann, L. W Molenkamp, X.-L. Qi, and S.-C. Zhang, 
Quantum Spin Hall Insulator State in HgTe Quantum Wells, 
Science {\bf 318}, 766 (2007). 

\bibitem{Hsieh2009}D. Hsieh, Y. Xia, D. Qian, L. Wray, J. H. Dil, F. Meier, J. Osterwalder, L. Patthey, J. G. Checkelsky, N. P. Ong, A. V. Fedorov, H. Lin, A. Bansil, D. Grauer, Y. S. Hor, R. J. Cava, and M. Z. Hasan, 
A tunable topological insulator in the spin helical Dirac transport regime, 
Nature (London) {\bf 460}, 1101 (2009). 

\bibitem{Hasan2010}M. Z. Hasan and C. L. Kane, 
Colloquium: Topological insulators, 
Rev. Mod. Phys. {\bf 82}, 3045 (2010). 


\bibitem{Xiao2011}D. Xiao, W. Zhu, Y. Ran, N. Nagaosa, and S. Okamoto, 
Interface engineering of quantum Hall effects in digital transition metal oxide heterostructures,
Nat. Commun. {\bf 2}, 596 (2011).


\bibitem{Ruegg2011}A. R{\"u}egg and G. A. Fiete, 
Topological insulators from complex orbital order in transition-metal oxides heterostructures, 
Phys. Rev. B {\bf 84}, 201103(R) (2011). 

\bibitem{Yang2011}K.-Y. Yang, W. Zhu, D. Xiao, S. Okamoto, Z. Wang, and Y. Ran, 
Possible interaction-driven topological phases in (111) bilayers of LaNiO$_3$,
Phys. Rev. B {\bf 84}, 201104(R) (2011).

\bibitem{Okamoto2014}
S. Okamoto, W. Zhu, Y. Nomura, R. Arita, D. Xiao, and N. Nagaosa,
Correlation effects in (111) bilayers of perovskite transition-metal oxides,
Phys. Rev. B {\bf 89}, 195121 (2014). 

\bibitem{Okamoto2018}S. Okamoto and D. Xiao,
Transition-metal oxide (111) bilayers, 
J. Phys. Soc. Jpn. {\bf 87}, 041006 (2018). 

\bibitem{Hirai2015}D. Hirai, J. Matsuno, and H. Takagi. 
Fabrication of (111)-oriented Ca$_{0.5}$Sr$_{0.5}$IrO$_3$/SrTiO$_3$ superlattices--A designed playground for honeycomb physics, 
APL Materials, {\bf 3}. 041508 (2015). 


\bibitem{Herring1937}C. Herring, Accidental degeneracy in the energy bands of crystals, Phys. Rev. {\bf 52}, 365 (1937). 

\bibitem{Wan2011}X. Wan, A. M. Turner, A. Vishwanath, and S. Y. Savrasov, 
Topological semimetal and Fermi-arc surface states in the electronic structure of pyrochlore iridates, 
Phys. Rev. B {\bf 83}, 205101 (2011). 

\bibitem{Burkov2011}A. A. Burkov, M. D. Hook, and L. Balents, 
Topological nodal semimetals, 
Phys. Rev. B {\bf 84}, 235126 (2011). 

\bibitem{Young2012}S. M. Young, S. Zaheer, J. C. Y. Teo, C. L. Kane, E. J. Mele, and A. M. Rappe, 
Dirac Semimetal in Three Dimensions, 
Phys. Rev. Lett. {\bf 108}, 140405 (2012). 

\bibitem{Young2015}S. M. Young and C. L. Kane, Dirac Semimetals in Two Dimensions, 
Phys. Rev. Lett. {\bf 115}, 126803 (2015). 

\bibitem{Burkov2016}A. A. Burkov, Topological semimetals, Nat. Mater. {\bf 15}, 1145 (2016). 

\bibitem{Bzdusek2016}T. Bzdu{\v s}ek, Q. Wu, A. R{\"u}egg, M. Sigrist, and A. A. Soluyanov, 
Nodal-chain metals, 
Nature (London) {\bf 538}, 75 (2016). 

\bibitem{Chiu2016}C.-K. Chiu, J. C. Y. Teo, A. P. Schnyder, and S. Ryu, 
Classification of topological quantum matter with symmetries, 
Rev. Mod. Phys. {\bf 88}, 035005 (2016). 

\bibitem{Armitage2018}N. P. Armitage, E. J. Mele, and A. Vishwanath, 
Weyl and Dirac semimetals in three-dimensional solids, 
Rev. Mod. Phys. {\bf 90}, 015001 (2018).

\bibitem{Burkov2018}A. A. Burkov, 
Weyl metals, 
Annu. Rev. Condens. Matter Phys. {\bf 9}, 359 (2018).

% Semi-Dirac

\bibitem{Pardo2009}V. Pardo and W. E. Pickett, 
Half-Metallic Semi-Dirac-Point Generated by Quantum Confinement in TiO$_2$/VO$_2$ Nanostructures,
Phys. Rev. Lett. {\bf 102}, 166803 (2009). 

\bibitem{Pardo2010}V. Pardo and W. E. Pickett, 
Metal-insulator transition through a semi-Dirac point in oxide nanostructures: VO$_2$ (001) layers confined within TiO$_2$, 
Phys. Rev. B {\bf 81}, 035111 (2010). 

\bibitem{Shibuya2010}K. Shibuya, M. Kawasaki, and Y. Tokura, 
Metal-insulator transitions in TiO$_2$/VO$_2$ superlattices
Phys. Rev. B {\bf 82}, 205118 (2010). 


\bibitem{Fujioka2019}J. Fujioka, R. Yamada, M. Kawamura, S. Sakai, M. Hirayama, R. Arita, T. Okawa, D. Hashizume, M. Hoshino, and Y. Tokura, 
Strong-correlation induced high-mobility electrons in Dirac semimetal of perovskite oxide, 
Nat. Commun. {\bf 10}, 362 (2019). 


\bibitem{Chen2015} Y. Chen, Y.-M. Lu, and H.-Y. Kee, 
Topological crystalline metal in orthorhombic perovskite iridates, 
Nat. Comm. {\bf 6}, 6593 (2015).

\bibitem{Ok2021} J. M. Ok, M. Mohanta, J. Zhang, S. Yoon, S. Okamoto, E. S. Choi, H. Zhou, M. Briggeman, P. Irvin, A. R. Lupini, Y.-Y. Pai, 
E. Skoropata, C. Sohn, H. Li, H. Miao, B. Lawrie, W. S. Choi, G. Eres, J. Levy, and H. N. Lee,
Correlated Oxide Dirac Semimetal in the Extreme Quantum Limit,
Sci. Adv. {\bf 7}, eabf9631 (2021).

\bibitem{Rosendal2023}
V. Rosendal, W. H. Brito, M. Radovic, A. Chikina, M. Brandbyge, N. Pryds, and D. H. Petersen, 
Octahedral distortions in SrNbO3: Unraveling the structure-property relation, 
Phys. Rev. Materials {\bf 7}, 075002 (2023). 

% Altermagnetism

\bibitem{Noda2016}Y. Noda, K. Ohnob, and S. Nakamura, 
Momentum-dependent band spin splitting in semiconducting MnO$_2$: a density functional calculation,
Phys. Chem. Chem. Phys. {\bf 18}, 13294 (2016). 

\bibitem{Okugawa2018}T. Okugawa, K. Ohno, Y. Noda, and S. Nakamura, 
Weakly spin-dependent band structures of antiferromagnetic perovskite LaMO$_3$ (M  =  Cr, Mn, Fe),
J. Phys.: Condens. Matter {\bf 30}, 075502 (2018). 


% THE in LSMO/SIO 

\bibitem{Mohanta2019}N. Mohanta, E. Dagotto, and S. Okamoto, 
Topological Hall effect and emergent skyrmion crystal at manganite-iridate oxide interfaces, 
Phys. Rev. B {\bf 100}, 064429 (2019). 

% Semi-Dirac in SrNbO3 

\bibitem{Mohanta2021}N. Mohanta, J. M. Ok, J. Zhang, H. Miao, E. Dagotto, H.-N. Lee, and S. Okamoto, 
Semi-Dirac and Weyl fermions in transition metal oxides, 
Phys. Rev. B {\bf 104}, 235121 (2021).


\bibitem{Slater1954}J. C. Slater and G. F. Koster, 
Simplified LCAO Method for the Periodic Potential Problem, 
Phys. Rev. {\bf 94}, 1498 (1954). 

% Berry phase

\bibitem{Xiao2010}D. Xiao, M.-C. Chang, and Q. Niu, Qian,
Berry phase effects on electronic properties, 
Rev. Mod. Phys. {\bf 82}, 1959 (2010).

\bibitem{Mohanta2020} N. Mohanta, A. D. Christianson, S. Okamoto, and E. Dagotto, 
Signatures of a liquid-crystal transition in spin-wave excitations of skyrmions, 
Commun. Phys. {\bf 3}, 229 (2020). 



% Elbio's section general framework 

\bibitem{elbio1-1} S. T. Bramwell and M. J. P. Gingras, Spin ice state in frustrated magnetic pyrochlore
materials, Science {\bf 294}, 1495 (2001).

\bibitem{elbio1-2} L. Balents, L. Spin liquids in frustrated magnets, Nature {\bf 464}, 199 (2010).

\bibitem{elbio1-3} L. Savary and L. Balents, Quantum spin liquids: a review, Rep. Prog. Phys. {\bf 80}, 016502
(2016).

\bibitem{elbio1-4} C. D. Batista, S.-Z. Lin, S. Hayami, and Y. Kamiya, Frustration and chiral orderings in
correlated electron systems, Rep. Prog. Phys. {\bf 79}, 084504 (2016).

\bibitem{elbio1} N. Mohanta and E. Dagotto, npj Quantum Mater. {\bf 7}, 76 (2022).

\bibitem{elbio1-5} Y. Taguchi, Y. Oohara, H. Yoshizawa, N. Nagaosa, and Y. Tokura, Spin chirality, Berry
phase, and anomalous Hall effect in a frustrated ferromagnet, Science {\bf 291}, 2573 (2001).

\bibitem{elbio1-6} N. Nagaosa and Y. Tokura, Topological properties and dynamics of magnetic skyrmions,
Nat. Nanotechnol. {\bf 8}, 899 (2013).

\bibitem{ramesh} S. Jin, T. H. Tiefeld, M. McCormack, R. A. Fastnacht, R. Ramesh, and I. H. Chen,
Thousandfold Change in Resistivity in Magnetoresistive La-Ca-Mn-O Films, Science {\bf 264}, 413 (1994).

% manganites phase diagram experiment 

\bibitem{Tokura1994}Y. Tokura, A. Urushibara, Y. Moritomo, T. Arima, A. Asamitsu, G. Kido, and N. Furukawa, 
Giant Magnetotransport Phenomena in Filling-Controlled Kondo Lattice System: La$_{1-x}$Sr$_x$MnO$_3$,
J. Phys. Soc. Jpn. {\bf 63}, 3931 (1994). 

\bibitem{Schiffer1995}P. Schiffer, A. P. Ramirez, W. Bao, and S-W. Cheong, 
Low Temperature Magnetoresistance and the Magnetic Phase Diagram of La$_{1-x}$⁢Ca$_x$⁢MnO3, 
Phys. Rev. Lett. {\bf 75}, 3336 (1995). 

\bibitem{Tokura1996}Y. Tokura, Y. Tomioka, H. Kuwahara, A. Asamitsu, Y. Moritomo, and M. Kasai, 
Origins of colossal magnetoresistance in perovskite‐type manganese oxides (invited), 
J. Appl. Phys. {\bf 79}, 5288 (1996). 

\bibitem{Tokura1999}Y. Tokura and Y. Tomioka, Colossal magnetoresistive manganites, J. Magn. Magn. Mater. {\bf 200} 1 (1999). 



\bibitem{elbio1-12} S. Yunoki, T. Hotta and E. Dagotto, Ferromagnetic, A-type, and charge-ordered CE type
states in doped manganites using Jahn–Teller phonons, Phys. Rev. Lett. {\bf 84}, 3714 (2000).

\bibitem{elbio1-13} A. Moreo, M. Mayr, A. Feiguin, S. Yunoki and E. Dagotto, Giant cluster coexistence in
doped manganites and other compounds, Phys. Rev. Lett. {\bf 84}, 5568 (2000).

\bibitem{elbio1-14} E. Dagotto, T. Hotta, and A. Moreo, Colossal magnetoresistant materials: the key
role of phase separation, Phys. Rep. {\bf 344}, 1 (2001).

\bibitem{elbio1-15} T. Hotta, M. Moraghebi, A. Feiguin, A. Moreo, S. Yunoki, and E. Dagotto, 
Unveiling new magnetic phases of undoped and doped manganites,
Phys. Rev. Lett. {\bf 90}, 247203 (2003).

\bibitem{hotta2000} T. Hotta, A. L. Malvezzi, and E. Dagotto, Charge-orbital ordering and phase separation in the two-orbital model for manganites: Roles of Jahn-Teller phononic and Coulombic interactions,  
Phys. Rev. B {\bf 62}, 9432 (2000). 


\bibitem{hotta2000b} T. Hotta, Y. Takada, H. Koizumi, and E. Dagotto,Topological Scenario for Stripe Formation in Manganese Oxides,  Phys. Rev. Lett. {\bf 84}, 2477 (2000). 

\bibitem{aliaga} H. Aliaga, D. Magnoux, A. Moreo, D. Poilblanc, S. Yunoki, and E. Dagotto, Theoretical Study of Half-Doped Models for Manganites: Fragility of the CE Phase with Disorder, Two Types of Colossal Magnetoresistances, and Charge-Ordered States for Electron-Doped Materials, Phys. Rev. B {\bf 68}, 104405 (2003).

\bibitem{elbio1-19} D. Yi,  J. Liu, S.-L. Hsu, L. Zhang, Y. Choi, J.-W. Kim, Z. Chen, J. D. Clarkson, C. R. Serrao, E. Arenholz, P. J. Ryan, H. Xu, R. J. Birgeneau, and R. Ramesh, 
Atomic-scale control of magnetic anisotropy via novel spin-orbit
coupling effect in La$_{2/3}$Sr$_{1/3}$MnO$_3$/SrIrO$_3$ superlattices, 
Proc. Natl. Acad. Sci. {\bf 113}, 6397 (2016).

%\bibitem{elbio1-20} E. Skoropata et al., Interfacial tuning of chiral magnetic interactions 
%for large topological Hall effects in LaMnO$_3$/SrIrO$_3$ heterostructures, Sci. Adv. {\bf 6}, 
%eaaz3902 (2020).

\bibitem{elbio1-7} A. Fert, N. Reyren, and V. Cros, Magnetic skyrmions: advances in physics and
potential applications, Nat. Rev. Mater. {\bf 2}, 17031 (2017).

\bibitem{elbio1-8} X. Z. Yu, Y. Onose, N. Kanazawa, J. H. Park, J. H. Han, Y. Matsui, N. Nagaosa, and Y. Tokura, 
Real-space observation of a two-dimensional skyrmion crystal,
Nature (London) {\bf 465}, 901 (2010).

\bibitem{elbio1-9} N. Mohanta, E. Dagotto and S. Okamoto, Topological Hall effect and emergent skyrmion
crystal at manganite-iridate oxide interfaces, Phys. Rev. B {\bf 100}, 064429 (2019).

\bibitem{elbio1-10} T. Okubo, S. Chung and H. Kawamura, Multiple-q states and the skyrmion lattice of
the triangular-lattice Heisenberg antiferromagnet under magnetic fields, Phys.
Rev. Lett. {\bf 108}, 017206 (2012).

\bibitem{elbio1-11} R. Ozawa, S. Hayami and Y. Motome, Zero-field skyrmions with a high topological
number in itinerant magnets, Phys. Rev. Lett. {\bf 118}, 147205 (2017).

\bibitem{elbio1-18} J. J. Nakane, K. Nakazawa and H. Kohno, Topological Hall effect in weakly canted
antiferromagnets, Phys. Rev. B {\bf 101}, 174432 (2020).

% specific example... 

\bibitem{elbio1-16} L.-C. Garnier, M. Marangolo, M. Eddrief, D. Bisero, S. Fin, F. Casoli, M. G. Pini, A. Rettori, and S, Tacchi,
Stripe domains reorientation in ferromagnetic films with
perpendicular magnetic anisotropy, J. Phys. Mater. {\bf 3}, 024001 (2020).

\bibitem{elbio1-17} C. B. Bishop, A. Moreo and E. Dagotto, Bicollinear antiferromagnetic order,
monoclinic distortion, and reversed resistivity anisotropy in FeTe as a result of
spin-lattice coupling, Phys. Rev. Lett. {\bf 117}, 117201 (2016).



%\bibitem{elbio2} N. Mohanta, E. Dagotto, and S. Okamoto, Topological Hall effect and emergent skyrmion
%crystal at manganite-iridate oxide interfaces, Phys. Rev. B {\bf 100}, 064429 (2019).

%\bibitem{elbio3} N. Mohanta, A. D. Christianson, S. Okamoto, and E. Dagotto, Signatures of a liquid crystal
%transition in spin-wave excitations of skyrmions. Commun. Phys. {\bf 3}, 229 (2020).

%\bibitem{elbio4} C. B. Bishop, A. Moreo, and E. Dagotto, Bicollinear antiferromagnetic order,
%monoclinic distortion, and reversed resistivity anisotropy in FeTe as a result of
%spin-lattice coupling, Phys. Rev. Lett. {\bf 117}, 117201 (2016).


\bibitem{Maezono1998a}R. Maezono, S. Ishihara, and N. Nagaosa, Orbital polarization in manganese oxides, 
Phys. Rev. B 57, R13993(R) (1998). 

\bibitem{Maezono1998b}R. Maezono, S. Ishihara, and N. Nagaosa, Phase diagram of manganese oxides, 
Phys. Rev. B 58, 11583 (1998). 

\bibitem{Okamoto2000}S. Okamoto, S. Ishihara, and S. Maekawa, Phase transition in perovskite manganites with orbital degree of freedom, 
Phys. Rev. B {\bf 61}, 14647 (2000). 

% Prediction of novel...

\bibitem{kancharla} S. Yunoki, A. Moreo, E. Dagotto, S. Okamoto, S. S. Kancharla, and A. Fujimori, 
Electron doping of cuprates via interfaces with manganites, Phys. Rev. B {\bf 76}, 064532 (2007).

\bibitem{canted-CE} R. Yu, S. Yunoki, S. Dong, and E. Dagotto, Phys. Rev. B {\bf 80}, 125115 (2009).


\bibitem{Goodenough1995}J. B. Goodenough, Theory of the Role of Covalence in the Perovskite-Type Manganites [La, $M$(II)]MnO$_3$, 
Phys. Rev. {\bf 100}, 564 (1955). 

% Metallicity 

\bibitem{LMO-SMO-elbio1} S. Dong, R. Yu, S. Yunoki, G. Alvarez, J.-M. Liu, and E. Dagotto, Phys. Rev. B {\bf 78}, 201102(R) (2008).

\bibitem{bhatta1} A. Bhattacharya, X. Zhai, M. Warusawithana, J. N. Eckstein, and
S. D. Bader, Appl. Phys. Lett. {\bf 90}, 222503 (2007).

\bibitem{bhatta2} S. Smadici, P. Abbamonte, A. Bhattacharya, X. Zhai, B. Jiang,
A. Rusydi, J. N. Eckstein, S. D. Bader, and J. M. Zuo, Phys.Rev. Lett. {\bf 99}, 196404 (2007).

\bibitem{bhatta3} S. J. May, A. B. Shah, S. G. E. te Velthuis, M. R. Fitzsimmons,
J. M. Zuo, X. Zhai, J. N. Eckstein, S. D. Bader, and A. Bhattacharya, Phys. Rev. B {\bf 77}, 174409 (2008).

\bibitem{bhatta4} A. Bhattacharya, S. J. May, S. G. E. te Velthuis, M. Warusawithana,
X. Zhai, B. Jiang, J. M. Zuo, M. R. Fitzsimmons, S. D. Bader, and J. N. Eckstein, Phys. Rev. Lett. {\bf 100}, 257203 (2008).



\bibitem{Okamoto2004b}S. Okamoto and A. J. Millis, Spatial inhomogeneity and strong correlation physics: A dynamical mean-field study of a model Mott-insulator—band-insulator heterostructure,
Phys. Rev. B {\bf 70}, 241104(R) (2004). 

% Quantumespresso

\bibitem{QS}P. Giannozzi, S. Baroni, N. Bonini, M. Calandra, R. Car, C. Cavazzoni, D. Ceresoli, G. L. Chiarotti, M. Cococcioni,
I. Dabo, A. Dal Corso, S. de Gironcoli, S. Fabris, G. Fratesi, R. Gebauer, U. Gerstmann, C. Gougoussis, A. Kokalj,
M. Lazzeri, L. Martin-Samos, N. Marzari, F. Mauri, R. Mazzarello, S. Paolini, A. Pasquarello, L. Paulatto, C. Sbraccia, 
S. Scandolo, G. Sclauzero, A. P. Seitsonen, A. Smogunov, P. Umari, and R. M. Wentzcovitch, 
QUANTUM ESPRESSO: a modular and open-source software project for quantum simulations of materials, 
J. Phys. Condens. Matter {\bf 21}, 395502 (2009).

% PBE 

\bibitem{PBE}J. P. Perdew, K. Burke, and M. Ernzerhof, 
Generalized gradient approximation made simple, 
Phys. Rev. Lett. {\bf 77}, 3865 (1996).

% Altermagnetism

%\bibitem{Noda2016}Y. Noda, K. Ohnob, and S. Nakamura, 
%Momentum-dependent band spin splitting in semiconducting MnO$_2$: a density functional calculation,
%Phys. Chem. Chem. Phys. {\bf 18}, 13294 (2016). 

%\bibitem{Okugawa2018}T. Okugawa, K. Ohno, Y. Noda, and S. Nakamura, 
%Weakly spin-dependent band structures of antiferromagnetic perovskite LaMO$_3$ (M  =  Cr, Mn, Fe),
%J. Phys.: Condens. Matter {\bf 30}, 075502 (2018). 

\bibitem{Naka2019}M. Naka, S. Hayami, H. Kusunose, Y. Yanagi, Y. Motome, and H. Seo, 
Spin current generation in organic antiferromagnets, 
Nat. Commun. {\bf 10}, 4305 (2019). 

\bibitem{Hayami2019}S. Hayami, Y. Yanagi, and H. Kusunose, 
Momentum-Dependent Spin Splitting by Collinear Antiferromagnetic Ordering, 
J. Phys. Soc. Jpn. {\bf 88}, 123702 (2019).

\bibitem{Hayami2020}S. Hayami, Y. Yanagi, and H. Kusunose, 
Bottom-up design of spin-split and reshaped electronic band structures in antiferromagnets without spin-orbit coupling: Procedure on the basis of augmented multipoles, 
Phys. Rev. B {\bf 102}, 144441 (2020).

\bibitem{Naka2021}M. Naka, Y. Motome, and H. Seo, 
Perovskite as a spin current generator, 
Phys. Rev. B {\bf 103}, 125114 (2021). 

\bibitem{Smejkal2022a}L. {\v S}mejkal, J. Sinova, and T. Jungwirth, 
Beyond Conventional Ferromagnetism and Antiferromagnetism: A Phase with Nonrelativistic Spin and Crystal Rotation Symmetry,
Phys. Rev. X {\bf 12}, 031042 (2022).

\bibitem{Smejkal2022b}L. {\v S}mejkal, J. Sinova, and T. Jungwirth, 
Emerging Research Landscape of Altermagnetism,
Phys. Rev. X {\bf 12}, 040501 (2022).

% 2nd order TSC 

\bibitem{Okamoto2013}S. Okamoto, Doped Mott insulators in (111) bilayers of perovskite transition-metal oxides with the strong spin-orbit coupling, Phys. Rev. Lett. {\bf 110}, 066403 (2013). 

\bibitem{Wang_PRB2011} F. Wang, and Y. Ran, Nearly flat band with chern number $C = 2$ on the dice lattice, Phys. Rev. B {\bf 84}, 241103 (2011).

\bibitem{Soni_PRB2020} R. Soni, N. Kaushal, S. Okamoto, and E. Dagotto, Flat bands and ferrimagnetic order in electronically correlated dice-lattice ribbons. Phys. Rev. B {\bf 102}, 045105 (2020).

\bibitem{Pfeiffer_PhysLett1969} E.R. Pfeiffer, and J.F. Schooley, Superconducting transition temperatures of Nb-doped SrTiO$_3$, Phys. Lett. A, {\bf 29}, 589, (1969) 

\bibitem{Binnig_PRL1980} G. Binnig, A. Baratoff, H. E. Hoenig, and J. G. Bednorz, Two-band superconductivity in Nb-Doped SrTiO$_3$, Phys. Rev. Lett. {\bf 45}, 1352  (1980). 

\bibitem{Dice_CommPhys2023} N. Mohanta, R. Soni, S. Okamoto and E. Dagotto, Majorana corner states on the dice lattice, Commun. Phys. {\bf 6}, 240 (2023).


\end{thebibliography}
\end{document}